\documentclass[12pt]{iopart}

\usepackage{graphicx}

\usepackage{enumitem}
\usepackage{siunitx}
\usepackage[hyperfootnotes=false]{hyperref}
\hypersetup{
    colorlinks=true,
    linkcolor=blue,
    filecolor=magenta,      
    urlcolor=cyan,
}
\usepackage{amsfonts}
\usepackage{nicefrac}
\usepackage{lipsum}
\usepackage{multirow}
\usepackage{color}
\usepackage{lineno}
\usepackage{xspace}
\usepackage{soul}
\usepackage{listing}
\usepackage{subfigure}

\usepackage{algorithm2e}
\usepackage{amsmath}
\usepackage{mathtools}
\usepackage{multirow}
\usepackage{listings} 
\usepackage{color}
\usepackage{here}
\usepackage{import}
\usepackage{bm}
\usepackage{adjustbox}

\renewcommand{\thefigure}{\Roman{figure}}

\suppressfloats

\makeatletter
\def\lst@makecaption{
  \def\@captype{table}
  \@makecaption
}

\newcommand{\mainmatter}{
  \setcounter{footnote}{0}
  \patchcmd{\@makefntext}{\fnsymbol}{\arabic}{}{}
  \patchcmd{\@thefnmark}{\fnsymbol}{\arabic}{}{}
  \def\@makefnmark{\textsuperscript{\arabic{footnote}}}
}
\makeatother

\definecolor{kw}{cmyk}{1, 0.50, 0, 0}

\begin{document}
\title[Fast Jet Tagging with MLP-Mixers on FPGAs]{Fast Jet Tagging with MLP-Mixers on FPGAs}

\author{Chang Sun\textsuperscript{1,*}, Jennifer Ngadiuba\textsuperscript{2}, Maurizio Pierini\textsuperscript{3}, Maria Spiropulu\textsuperscript{1}}
\address{\textsuperscript{1}California Institute of Technology, Pasadena, CA, USA}
\address{\textsuperscript{2}Fermi National Accelerator Laboratory, Batavia, IL, USA}
\address{\textsuperscript{3}European Organization for Nuclear Research (CERN), Geneva, Switzerland}

\ead{chsun@cern.ch}

\vspace{10pt}

\thispagestyle{plain}
\pagestyle{plain}

\begin{abstract}
    We explore the innovative use of MLP-Mixer models for real-time jet tagging and establish their feasibility on resource-constrained hardware like FPGAs. MLP-Mixers excel in processing sequences of jet constituents, achieving state-of-the-art performance on datasets mimicking Large Hadron Collider conditions. By using advanced optimization techniques such as High-Granularity Quantization and Distributed Arithmetic, we achieve unprecedented efficiency. These models match or surpass the accuracy of previous architectures, reduce hardware resource usage by up to 97\%, double the throughput, and half the latency. Additionally, non-permutation-invariant architectures enable smart feature prioritization and efficient FPGA deployment, setting a new benchmark for machine learning in real-time data processing at particle colliders.

\end{abstract}

\mainmatter
\section{Introduction}
\label{sec:introduction}

The CERN Large Hadron Collider (LHC)~\cite{lhc1995large} generates vast quantities of data, producing hundreds of terabytes per second from proton-proton collisions occurring every 25 ns. The first reduction of this enormous data throughput is managed by the hardware-based level-1 trigger system (L1T), which filters events in real-time with stringent latency requirements of a few microseconds~\cite{cms-tdr-021,atlas-tdr-029}. This system, utilizing approximately 1,000 Field-Programmable Gate Arrays (FPGAs), ensures critical events are preserved for offline analysis while drastically reducing the data rate by two orders of magnitude, thereby managing the downstream bandwidth effectively. In this context, the algorithms' accuracy is crucial while the computational complexity of each algorithm must be kept minimal as the resources are scarce. For each individual algorithm, the latency requirement could range from a few microseconds down to a few hundred nanoseconds depending on the specific task. A throughput of 40 MHz, the rate at which the L1T receives data, is required for all algorithms.
The forthcoming High-Luminosity LHC (HL-LHC)~\cite{hl-lhc} upgrade will significantly increase the data rate and complexity, posing challenges for real-time data processing. Machine learning techniques are being actively explored to enhance the speed and accuracy of the trigger algorithms~\cite{cms-tdr-021,atlas-tdr-029}; however, addressing strict resource and latency constraints without compromising performance present important challenges.

Jet tagging is a crucial task for physics analyses at the LHC,
which enables the identification of rare events that probe fundamental
questions about the universe, such as the existence of new particles or
the verification of theoretical models. This task involves identifying the origin of jets, which are collimated sprays of particles produced in high-energy collisions, and associating them with specific underlying physics processes. While jet tagging is traditionally performed offline using large machine learning models and high-performance computing resources~\cite{pnet, part, pelican, lagtr, jedi-net}, extending this capability to the L1T ensures that such valuable events are not discarded prematurely. However, this transition is highly challenging due to the stringent constraints of the trigger system. Offline models, designed for accuracy without hardware constraints, often exceed the memory, processing speed, and on-chip resource availability of FPGAs, necessitating significant architectural adaptations and optimizations to make them viable for real-time deployment.

MLP-Mixer~\cite{mlp-mixer} is an architecture consisting solely of Multi-layer Perception (MLP) layers, originally designed for vision tasks. For those applications, it splits the input into patches and processes them as a sequence of vectors. As the inputs for jet-tagging tasks at the trigger level are already sequences of particles constituents, MLP-Mixer offers a plausible structure for online jet tagging tasks. In this work, we explore and demonstrate the application of MLP-Mixers for jet-tagging, as well as their deployment on FPGAs for use in low-level trigger systems. We train the models with High-Granularity Quantization (HGQ)~\cite{hgq} to fit within FPGA constraints, and further optimizes their firmware with Distributed Arithmetic (DA) to minimize the latency and maximizing throughput. The firmware is generated and deployed using the \texttt{hls4ml}~\cite{hls4ml} framework with the Vitis HLS~\cite{vitis} backend, ensuring compatibility and efficiency on FPGA hardware.

This work makes the following contributions:
\begin{itemize}[noitemsep,topsep=0pt]
    \item We show that MLP-Mixer models achieve state-of-the-art accuracy on a jet tagging dataset generated to closely resemble realistic LHC data.
    \item We demonstrate the deployment of MLP-Mixer models on FPGAs, achieving low latency and high throughput, meeting the requirements of the L1T system at the HL-LHC.
    \item We highlight that MLP-Mixer models outperform prior architectures in accuracy while using significantly fewer resources, achieving higher throughput, and lower latency on FPGAs.
    \item We emphasize the advantage of using a non-permutation-invariant model for jet tagging when leveraging HGQ, which allows for particle-wise selective feature prioritization via heterogeneous activation bitwidths to minimize resource consumption.
\end{itemize}

\section{Background and Related Works}
\label{sec:relatedwork}

\subsection{Fast Jet tagging on FPGAs}
Jet tagging is one of the most critical and well-studied tasks in high-energy physics experiments, with various machine learning algorithms proposed to address it~\cite{pnet,part,pelican,lagtr,jedi-net,efn}. State-of-the-art models for jet tagging include those based on Graph Neural Networks (GNN), such as Particle Net~\cite{pnet} and PELLICAN~\cite{pelican}, or Transformer architectures like Particle Transformer~\cite{part} and the Lorentz-Equivariant Geometric Algebra Transformers~\cite{lagtr}.

While these models demonstrate excellent performance for offline data processing, they are typically too large and computationally intensive for real-time deployment on FPGA-based trigger systems. Research efforts~\cite{tgc,qkeras,qkeras-xtre-q,autoq,hls4ml,llp,compress-pipe} have focused on adapting machine learning models for FPGA deployment in real-time environments, showcasing their potential for ultra-low latency and high throughput inference. Notably, Ref.~\cite{qkeras} demonstrated that quantization-aware training (QAT) could enable jet tagging with quantized neural networks on FPGAs, achieving competitive accuracy to the full precision models with significantly reduced resource usage. However, these efforts often relied on toy datasets with high-level jet features which are unavailable in real-time trigger systems, making them impractical for actual deployment.

For real-time jet tagging on FPGAs, models like Particle Net and Particle Transformer are quickly ruled out due to their size and complexity. More resource-efficient alternatives include JEDI-net~\cite{jedi-net}, an efficient architecture based on Interaction Networks (IN)~\cite{in}, and models using the Deep Sets (DS) architecture~\cite{ds}, are being explored. JEDI-net has been successfully implemented on FPGAs~\cite{jedi-fpga} with realistic latency and resource consumption, while the DS-based model achieved better latency and resource utilization than IN-based models under comparable hardware-aware constraints~\cite{ds-fpga}. These advancements highlight the feasibility of deploying highly-performant particle-based machine learning models on FPGAs for jet tagging, but further innovations are needed to balance performance, resource efficiency, and latency.

\subsection{hls4ml}
\texttt{hls4ml}~\cite{hls4ml} is a framework that translates trained neural networks into \texttt{C++} based high-level synthesis (HLS) projects, which can be further compiled by HLS backends into description languages such as Verilog or VHDL. It supports a wide range of neural network architectures and HLS backends. When the input neural network is properly quantized and configured, \texttt{hls4ml} may generate firmware that is bit-accurate with the original model. This framework has been widely used for deploying neural networks on the hardware triggers at the LHC~\cite{qkeras,qkeras-xtre-q,tgc,ae-l1t,llpnet,axol1tl,cicada,cms-tdr-021,atlas-tdr-029}. Notably, \texttt{hls4ml} has been employed for implementing autoencoder-based anomaly detection triggers~\cite{axol1tl,cicada} in the L1T system of the CMS experiment, demonstrating its reliability and effectiveness on FPGAs for physics-based applications and achieving remarkable success during LHC Run-3 data taking.

In this work, we leverage \texttt{hls4ml} with the HGQ~\cite{hgq} front-end and the Vitis HLS backend to generate the firmware for the MLP-Mixer models. This combination ensures bit-accuracy between the quantized python model and the generated firmware, while optimizing for resource efficiency on FPGAs.

\subsection{MLP-Mixer}
MLP-Mixer~\cite{mlp-mixer} is a neural network architecture that exclusively uses Multi-layer Perception (MLP) layers, which is originally designed for vision tasks. For a vision task, the model first splits the image into patches, and treat the flattened vector from each patch as the feature embeddings.
The core architecture of the networks involves a series of MLP layers applied along orthogonal directions to mix both spatial- and channel-wise information. Despite its simplicity, this approach has shown competitive performance in computer vision tasks. Since the input to the jet tagging model is already a sequence of vectors of particle features at L1 triggers, we directly apply the MLP-Mixer architecture on this data without additional modifications. The absence of complex mechanisms, such as self-attention in Transformers involving multiplication of variable-to-variable multiplication operations, simplifies the conversion of MLP-Mixer models with \texttt{hls4ml} and therefore their deployment on FPGAs and its optimization with DA.

\section{Setup}
\label{sec:setup}
\subsection{Dataset}
\label{sec:dataset}
\begin{figure*}[htb]
    \centering
    \subfigure[]{
        \includegraphics[width=0.4\textwidth]{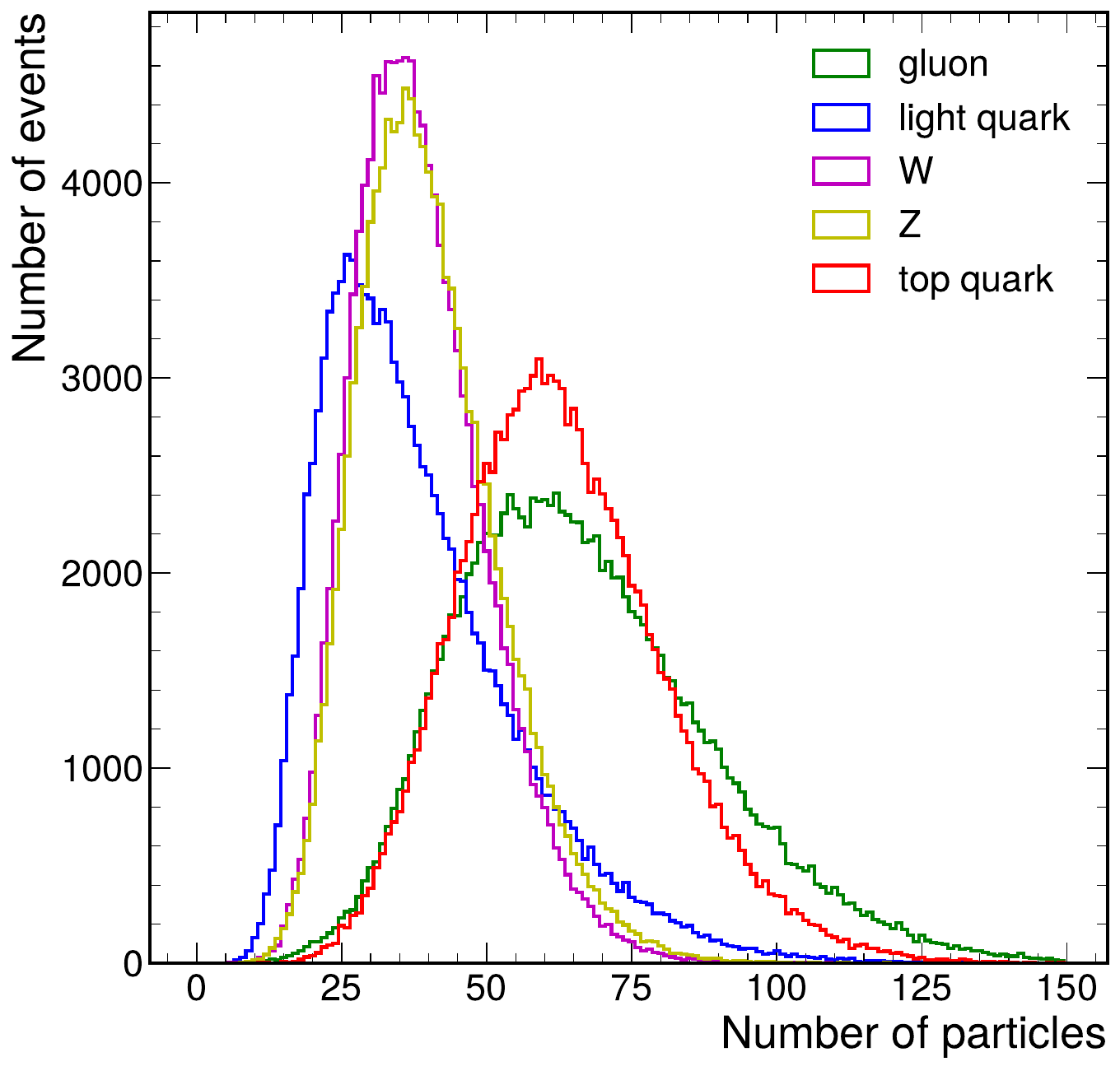}
        \label{fig:nptl:1}
    }
    \subfigure[]{
        \includegraphics[width=0.4\textwidth]{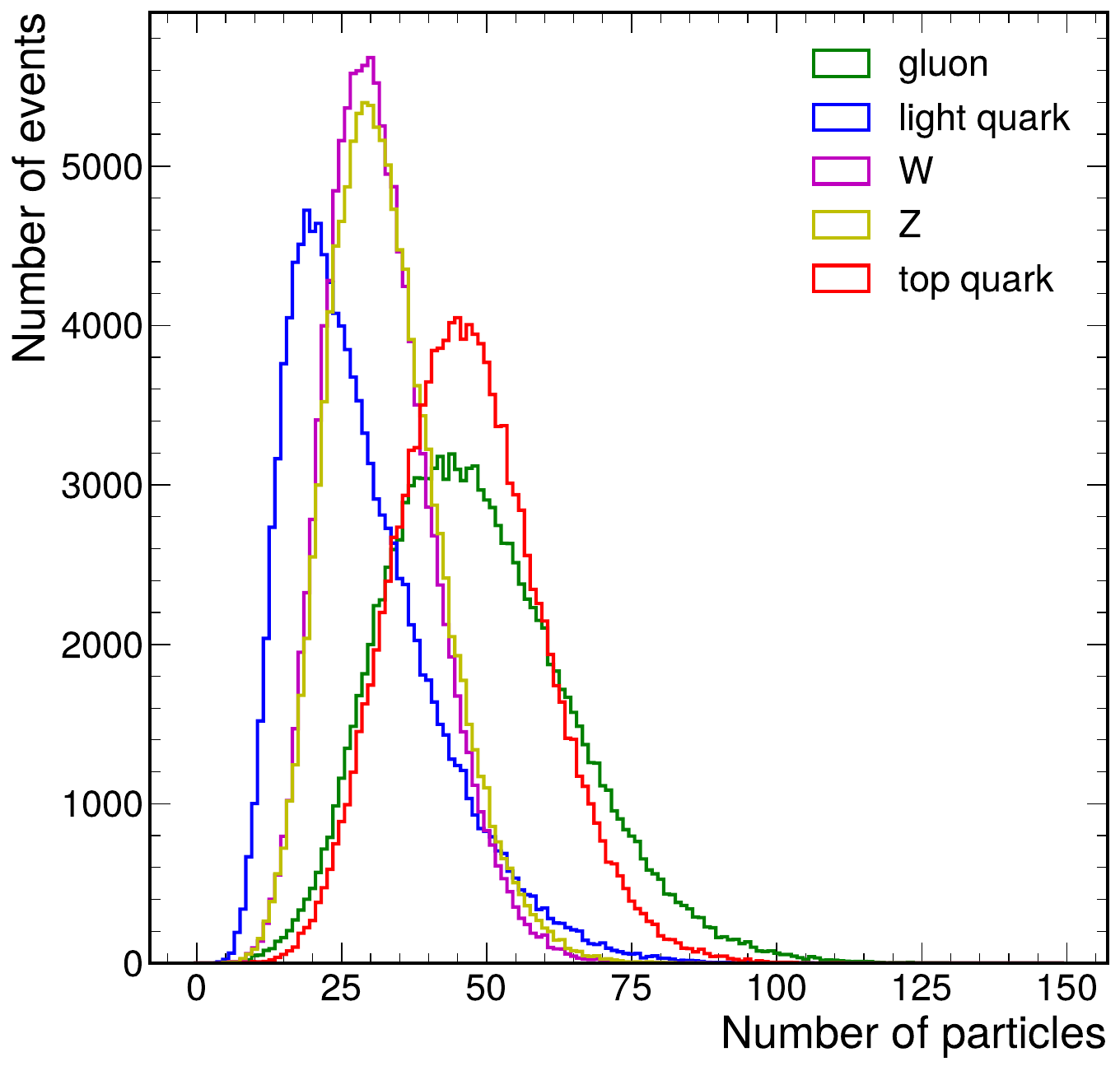}
        \label{fig:nptl:2}
    }
    \caption{Distribution of the number of particles per jet in the jet tagging dataset. On the right, only particles with $p_T \ge 2$ GeV are considered.}
\end{figure*}

We use the hls4ml jet tagging dataset~\cite{hls4ml-dataset} for training and evaluation. The dataset contains simulated jets originated from p-p collisions at $\sqrt{s}=13$ TeV. Five classes of jets are included: Jets originating from gluons (g), light quarks (q), W bosons (W), Z bosons (Z), and top quarks (t). All these jets are originated from di-jet events, where both patron or undecayed gauge bosons are generated with $p_T \approx 1$ TeV. Together with the underlying event, the particles are clustered with the AK8 algorithm. This dataset comprises 620,000 jets in the training set and 260,000 jets in the validation set, both balanced across the five classes. Each jet contains up to 150 particles, and each particle is characterized by 16 kinematic features, as documented in Table~\ref{tab:features}. The distributions of particle count for each class are shown in Figure~\ref{fig:nptl:1}, with the average number of particles per jet varying from 38 to 67, depending on the class.

\begin{table*}[htb]
    \centering
    \caption{Description of the 16 kinematic features used for each particle in the hls4ml jet tagging dataset. For the exact definition of each feature, please refer to~\cite{hls4ml-dataset}.}
    \label{tab:features}
    \begin{tabular}{|l|l|}
        \hline
        Feature                   & Description                                                                                                  \\
        \hline
        $p_x$                     & The $x$ component of the momentum of the particle.                                                           \\
        $p_y$                     & The $y$ component of the momentum of the particle.                                                           \\
        $p_z$                     & The $z$ component of the momentum of the particle.                                                           \\
        $E$                       & The energy of the particle.                                                                                  \\
        $E_\mathrm{rel}$          & $E/E_\mathrm{jet}$, relative energy of the particle.                                                         \\
        $p_T$                     & Transverse momentum of the particle.                                                                         \\
        ${p_T}_\mathrm{rel}$      & $p_T/{p_T}_\mathrm{jet}$, relative transverse momentum of the particle.                                      \\
        $\eta$                    & Pseudorapidity of the particle.                                                                              \\
        $\eta_\mathrm{rel}$       & $\eta-\eta_\mathrm{jet}$, relative pseudorapidity of the particle.                                           \\
        $\eta_\mathrm{rot}$       & $\eta$ rotated as described in~\cite{rot}                                                                    \\
        $\phi$                    & Azimuthal angle of the particle.                                                                             \\
        $\phi_\mathrm{rel}$       & $\phi-\phi_\mathrm{jet}$, relative azimuthal angle of the particle.                                          \\
        $\phi_\mathrm{rot}$       & $\phi$ rotated as described in~\cite{rot}                                                                    \\
        $\Delta R$                & $\sqrt{(\eta-\eta_\mathrm{jet})^2+(\phi-\phi_\mathrm{jet})^2}$, $\Delta R$ between the particle and the jet. \\
        $\cos\theta$              & $\cos(\theta)$, cosine of the polar angle of the particle.                                                   \\
        $\cos\theta_\mathrm{rel}$ & $\cos(\theta - \theta_\mathrm{jet})$, the cosine of relative polar angle of the particle.                    \\
        \hline
    \end{tabular}
\end{table*}

In alignment with trigger system constraints, only the top-$N$ particles ordered by $p_T$ are considered for real-time jet tagging tasks, with $N$ varying up to a realistic constraint.

Furthermore, one may apply a typical preselection requirement of a minimum transverse momentum ($p_T$) of 2 GeV per particle to simulate the constraints of the trigger system at HL-LHC~\cite{SCRecoL1Jet}.
With this $p_T$ threshold, the average particle count per jet decreases to a range of 31 to 49 as shown in Figure~\ref{fig:nptl:2}. If a jet contains fewer than $N$ particles, we pad the missing features with zeros.

In this study, we use 558,000 jets from the training set for model training, reserving the remaining 62,000 jets for validation. The remaining 260,000 jets from the validation set are used exclusively for testing.

Currently, the JEDI-net models are considered the state-of-the-art for this dataset, and we use them as the baseline for comparison in our study.

At the CMS Experiment~\cite{cms-tdr-021}, we expect the information of up to 128 particles to be available in the trigger system~\cite{SCRecoL1Jet} after the phase-2 upgrade. However, the number of particles required for downstream tasks, varies depending on the application. For instance, while it has been shown that a good performance in b-tagging requires only $\sim10$ particles~\cite{L1BTag}, we have observed that the jet-tagging task on this dataset benefits from a larger number of particles in the input. Also, event-level tasks, such as missing transverse energy estimation or anomaly detection, may benefit from a larger input to represent the full event. To account for these variations, we train the models with 16, 32, 64, and 128 particles to cover the expected range of inputs.

\subsection{Model}

\begin{figure}[htb]
    \centering
    \includegraphics[width=0.8\textwidth]{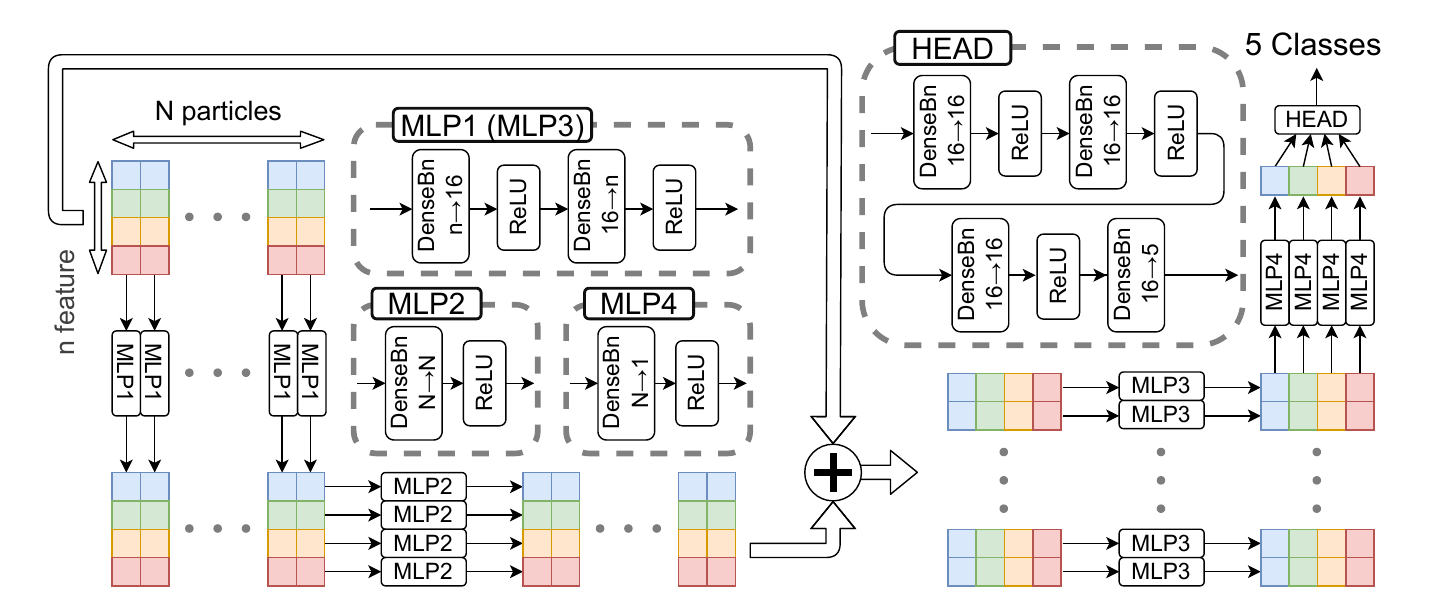}
    \caption{Architecture of the MLP-Mixer models used in this work. Each model consists of four MLP blocks with a single skip-connection. The input channel size match the number of per-particle input features. The implementation of each MLP and the classification head is shown in the corresponding dashed blocks. MLP1 and MLP3 act on the feature dimension; MLP2 and MLP4 act on the particle dimension. \texttt{DenseBn} represents a dense layer followed by a batch normalization layer during training, which are fused into a single layer during inference. The exact implementation of the network can be found in the in the repository in Section~\ref{sec:data_avail}.}
    \label{fig:model}
\end{figure}

We adopt the MLP-Mixer architecture proposed in~\cite{mlp-mixer} as the backbone of our model. However, since the original model is too large for practical FPGA implementation, we designed a smaller version with only two mixer stages and a dense classification head. Nevertheless, we find that this simplified model is still capable to achieve state-of-art-performance. The exact architecture is depicted in Figure~\ref{fig:model}. Following~\cite{jedi-fpga}, we use all 16 particle features available in the dataset, and train the models with varying number of particles: 16, 32, 64, and 128.

As a baseline for comparison, we also train an MLP model with three hidden layers of size 64 with the same number of particles in the input. The MLP model is trained and synthesized with the same configuration and optimizations as the MLP-Mixer models.

In contrast to \cite{jedi-fpga}, our models are not permutation-invariant, and all inputs are ordered by $p_T$. This design choice is made for several reasons:
\begin{itemize}[noitemsep,topsep=0pt]
    \item In practical trigger systems, particles are often already sorted by $p_T$, to meet the requirements of parallel algorithms running on the same FPGA.
    \item Allowing the model to break permutation invariance enables it to learn which features to retain and which to discard, facilitating heterogeneous bitwidths optimization on its activations.
    \item While permutation invariance is ensured in previous works on the top-$N$ elements, a preprocessing step is still required to extract the top-$N$ elements by $p_T$. Depending on the exact implementation, the resource overhead of sorting the whole sequence and taking the top-$N$ elements are likely to be comparable.
\end{itemize}

\subsection{Training}
All models are implemented and trained with \texttt{TensorFLow 2.13}. We use the Adam optimizer with an initial learning rate of $5\cdot10^{-3}$, along with the \texttt{CosineDecayRestarts} scheduler that resets the learning rate every 100 epochs. The full-precision models are trained for 500 epochs with a batch size of 512 with the categorical cross-entropy loss, and all models are trained on a single NVIDIA GTX 1080 GPU with 8GB of memory. We use the model with the best validation accuracy for performance evaluation described in Section~\ref{sec:full_precision} for the full-precision models.

\section{Quantization and Compression}

Floating-point representations are commonly used for weights and activations in neural networks. However, such representations are not suitable for FPGA implementation due to their high computational complexity of multiplication and accumulation operations. To deploy models efficiently on FPGAs, we quantize all neural network parameters to fixed-point representations.

\begin{figure}[htb]
    \centering
    \includegraphics[width=0.8\textwidth]{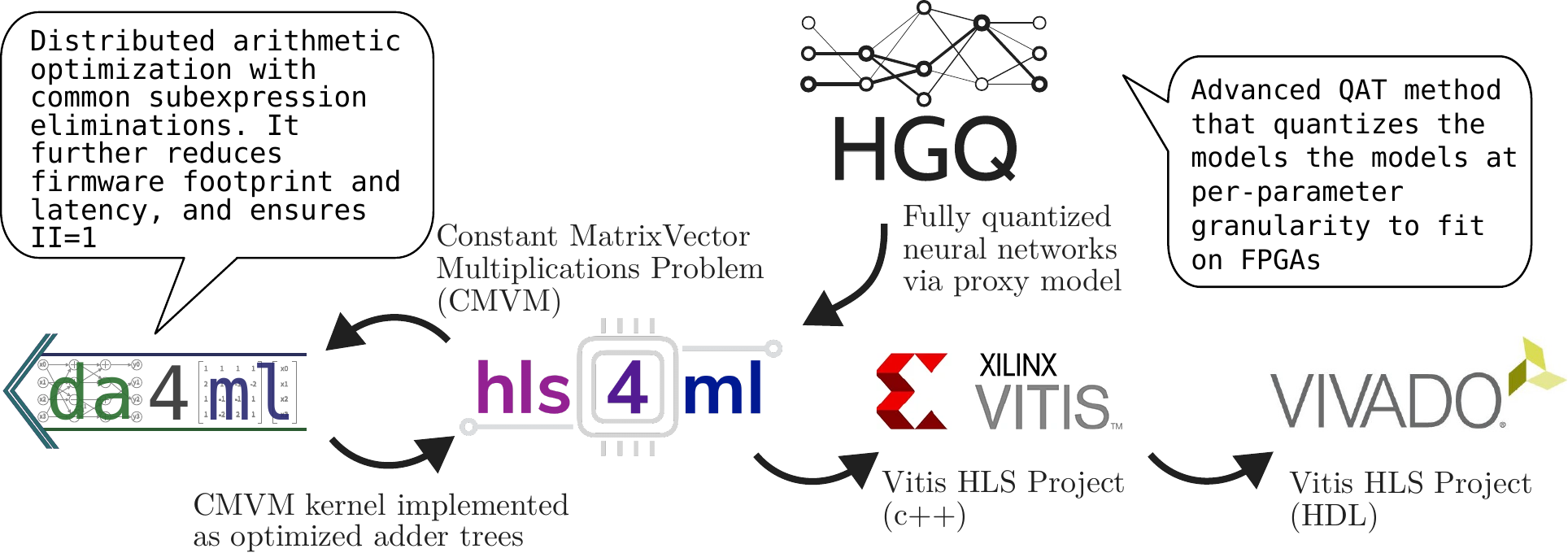}
    \caption{Workflow for training and deploying the MLP-Mixer models.}
    \label{fig:workflow}
\end{figure}

We adopt unified QAT and pruning provided by HGQ to make the model more compact and optimized for hardware deployment. Firmware generation is performed using \texttt{hls4ml} and Vitis HLS. DA optimization is performed using \texttt{da4ml} during firmware generation. The complete workflow for generating FPGA firmware is illustrated in Figure~\ref{fig:workflow}.

\subsection{Optimization Techniques}
We adopt the following techniques to optimize the model for FPGA deployment.

\subsubsection{High Granularity Quantization}
We adopt the HGQ~\cite{hgq} method to train, quantize, and prune the MLP-Mixer models. HGQ is an advanced QAT method that optimizes the bitwidths of each individual parameter of the neural network with surrogate gradients on bitwidths. Comparing to traditional QAT methods that optimize the bitwidths at the layer level, such as QKeras~\cite{qkeras} or Brevitas~\cite{brevitas}, HGQ provides a more fine-grained optimization that can achieve significant resource usage reduction for a fully unrolled firmware design while maintaining the accuracy of the original full-precision model. As HGQ allow the bitwidths to reach zero, unstructured pruning is naturally provided. HGQ is a crucial component of our workflow, which ensures that the MLP-Mixer models can be successfully deployed on FPGAs, as synthesis becomes infeasible if the model is not sufficiently quantized.

\subsubsection{Distributed Arithmetic}

DA~\cite{hcmvm, two-term, num-split, mmcm, fir1} is a method that transforms a linear digital signal processing (DSP) problem with one constant term, such as finite impulse response filters, discrete sine transform, or discrete cosine transform, into a series of operations implemented without the need for dedicated multipliers. This typically involves decomposing constant multiplication operations into a sequence of additions/subtractions and bit-shifts. Optimizations can be applied to the DA implementation to reduce the number of operations and the resource usage. In this work, we adopt a variant of the constant matrix-vector multiplication (CMVM) algorithm proposed in \cite{two-term} to optimize the CMVM operations in the MLP-Mixer models. An outline of the optimization is as follows:
\begin{enumerate}
    \item Decompose all constants in the operand matrix using the canonical signed digit representation, and express the CMVM operation as a sequence of additions/subtractions and bit-shifts.
    \item Iteratively search for repeated two-term subexpressions and replace them with a single variable.
    \item Trace the necessary bitwidths for all intermediate variables and the final results to prevent overflows.
\end{enumerate}
We use the \texttt{da4ml}~\cite{da4ml-repo} library as an extension to the \texttt{hls4ml} library to apply this optimization. By applying the stated DA optimization, we significantly reduce the latency and initiation interval of the models.

\subsubsection{Fused Batch Normalization}
Since a dense layer performs a general affine transformation and a batch normalization (BN) layer applies a more restricted affine transformation, a BN layer immediately following a dense layer can be fused with the latter by scaling the kernel and shifting the bias accordingly. However, because the weights of the dense layer are expected to be highly quantized, post-training fusion may lead to suboptimal accuracy and/or resource consumption overheads. To address this, we use the fused \texttt{DenseBatchnorm} layer from the HGQ library, which fuses the BN layers into the dense layer during training. This ensures that quantization bitwidths are optimized for the fused layers. As a result, both bitwidths and model weights are tuned appropriately, preventing any degradation in performance or resource efficiency.

\subsection{Training}

Following~\cite{hgq}, we use a single training for each model to cover the entire Pareto frontier between accuracy and resource consumption\footnote{The set of models that achieve the best accuracy for a given resource consumption, or the lowest resource consumption for a given accuracy.} for the quantized models. We initialize the models with a small $\beta$ -- a scalar parameter that encourages smaller models in a hardware resource-aware approach though regularization on differentiable bitwidths -- and increase its value during training with an exponential scheduler. In each training step, the approximated Effective Bit-Operations ($\overline{\mbox{EBOPs}}$)~\cite{hgq} -- a differentiable resource surrogate when deployed on FPGAs for the model computed from the bitwidths from its parameters -- is scaled by $\beta$ and added to the loss. In this way, $\beta$ leads to lower bitwidths, and consequently lower resource utilization on FPGAs. More details on $\beta$ and $\overline{\mbox{EBOPs}}$ can be found in~\cite{hgq}. During training, we save all models that lie on the Pareto frontier between validation accuracy and the resource consumption estimated by $\overline{\mbox{EBOPs}}$. All other hyperparameters remain identical to those used in the full-precision training. Each model is trained for 5,000 epochs to cover the Pareto frontier. However, the training time can be significantly reduced if only one performing model is required. For instance, all best-performing quantized models depicted in this work are obtained within 400 epochs.

\section{Results}

In this section, we first evaluate the performance of the full-precision MLP-Mixer models on the jet tagging dataset and compare them with the JEDI-net models~\cite{jedi-net} using different numbers of input particles. We then analyse the performance, resource consumption, latency, and throughput of the deployable firmware for both MLP-Mixer and MLP models on the same dataset. Additionally, we compare quantized models with their full-precision counterparts and with previous studies to highlight the advantages of using MLP-Mixer-based architectures for this task. For fairness, the $p_T\ge2$ GeV selection was not applied to the input particles when comparing our results with~\cite{jedi-fpga} and~\cite{jedi-net}.

Finally, we examine the features selected by the MLP-Mixer models and demonstrate the benefits of employing non-permutation-invariant architectures for real-time jet tagging.

\subsection{Full Precision Model}
\label{sec:full_precision}

The performance of the full-precision MLP-Mixer models (MLPM-fp) and JEDI-net~\cite{jedi-net} models is summarized in Table~\ref{tab:full}. The four MLP-Mixer models take 16, 32, 64, and 128 particles as input, while the two most accurate JEDI-net variants use 100 and 150 input particles. To represent the input configuration, we use ${N_p}^{\mathrm{\#fea}}=\mathrm{\#ptl}$, where $\mathrm{\#fea}$ denotes the number of features per particle and $\mathrm{\#ptl}$ represents the maximum number of particles used as input. Both MLP-Mixer and JEDI-net models utilize all 16 particle features, as described in Section~\ref{sec:dataset}. For MLP-Mixer models, the top-$N$ particles are selected and ordered by $p_T$ before being fed into the model. In contrast, the JEDI-net models also take as input the top-$N$ particles by $p_T$ but do not rely on their specific order, as they utilize a permutation-invariant architecture.

The performance metrics, including the area-under-curve (AUC) and the true positive rate (TPR) at false positive rate (FPR) of $1\%$ and $10\%$, are presented in Table~\ref{tab:full} for each class, with the best performance in each category highlighted in bold. The MLP-Mixer models consistently achieve higher AUC scores than the JEDI-net models across all classes. However, in terms of TPR at FPR=$1\%$ and FPR=$10\%$, their performance varies slightly, with some classes showing marginal improvements while others experience minor reductions. Among all MLP-Mixer variants, the model with 64 input particles delivers the best overall performance across all categories.

\begin{table*}[htb]
    \centering
    \def\jedi{ (${N_p}^{16}$=100)~\cite{jedi-net}}
    \def\jedio{JEDI-net w/ $\Sigma\mathcal{O}$ (${N_p}^{16}$=150)~\cite{jedi-net}}
    \def\so{$\Sigma\mathcal{O}$}
    \caption{Performance comparison of the MLP-Mixer and JEDI-net models. The table reports the AUC and the TPR at FPR of $1\%$ and $10\%$, for the different jet categories. The MLP-Mixer models are trained with 16, 32, 64, and 128 input particles, while the JEDI-net models use 100 and 150 input particles, where each particle is described by 16 features. The best performance in each category is highlighted in bold.}
    \label{tab:full}
    \begin{adjustbox}{width=1.0\textwidth,center=\textwidth}
        \begin{tabular}{l|cccccc}
            \hline
            \multirow{2}{*}{Model} & JEDI-net~\cite{jedi-net}              & JEDI-net               & MLPM-fp          & MLPM-fp          & MLPM-fp               & MLPM-fp          \\
                                   &                                       & w/ \so~\cite{jedi-net} &                  &                  &                       &                  \\
                                   & ${N_p}^{16}=100$                      & ${N_p}^{16}=150$       & ${N_p}^{16}=16$  & ${N_p}^{16}=32$  & ${N_p}^{16}=64$       & ${N_p}^{16}=128$ \\
            \hline
                                   & \multicolumn{6}{c}{AUC (\%)}                                                                                                                    \\
            \hline

            gluon                  & $95.29 \pm 0.01$                      & $95.28 \pm 0.01$       & $94.08 \pm 0.03$ & $95.08 \pm 0.04$ & $\bm{95.53 \pm 0.05}$ & $95.48 \pm 0.05$ \\
            light quarks           & $93.01 \pm 0.01$                      & $92.90 \pm 0.01$       & $92.15 \pm 0.03$ & $93.09 \pm 0.04$ & $\bm{93.43 \pm 0.05}$ & $93.32 \pm 0.03$ \\
            W boson                & $97.39 \pm 0.01$                      & $96.95 \pm 0.01$       & $96.06 \pm 0.02$ & $97.49 \pm 0.03$ & $\bm{97.83 \pm 0.02}$ & $97.78 \pm 0.03$ \\
            Z boson                & $96.79 \pm 0.01$                      & $96.49 \pm 0.01$       & $95.34 \pm 0.02$ & $97.12 \pm 0.03$ & $\bm{97.50 \pm 0.04}$ & $97.43 \pm 0.03$ \\
            top quark              & $96.83 \pm 0.01$                      & $96.77 \pm 0.01$       & $96.14 \pm 0.03$ & $96.91 \pm 0.02$ & $\bm{97.13 \pm 0.02}$ & $97.01 \pm 0.03$ \\

            \hline
                                   & \multicolumn{6}{c}{TPR at FPR=$10\%$}                                                                                                           \\
            \hline

            gluon                  & $87.8 \pm 0.1$                        & $\bm{87.9 \pm 0.1}$    & $82.2 \pm 0.1$   & $85.3 \pm 0.2$   & $86.7 \pm 0.2$        & $86.6 \pm 0.1$   \\
            light quarks           & $\bm{82.2 \pm 0.1}$                   & $81.8 \pm 0.1$         & $77.8 \pm 0.1$   & $80.3 \pm 0.1$   & $81.1 \pm 0.1$        & $81.0 \pm 0.1$   \\
            W boson                & $\bm{93.8 \pm 0.1}$                   & $92.7 \pm 0.1$         & $89.5 \pm 0.1$   & $93.1 \pm 0.1$   & $\bm{93.8 \pm 0.1}$   & $93.7 \pm 0.1$   \\
            Z boson                & $91.0 \pm 0.1$                        & $90.3 \pm 0.1$         & $86.4 \pm 0.1$   & $91.2 \pm 0.1$   & $\bm{92.2 \pm 0.1}$   & $92.0 \pm 0.1$   \\
            top quark              & $93.0 \pm 0.1$                        & $\bm{93.1 \pm 0.1}$    & $90.8 \pm 0.1$   & $92.5 \pm 0.1$   & $93.0 \pm 0.1$        & $92.7 \pm 0.1$   \\

            \hline
                                   & \multicolumn{6}{c}{TPR at FPR=$1\%$}                                                                                                            \\
            \hline

            gluon                  & $\bm{48.5 \pm 0.1}$                   & $48.2 \pm 0.1$         & $40.8 \pm 0.2$   & $44.9 \pm 0.3$   & $46.9 \pm 0.3$        & $46.6 \pm 0.5$   \\
            light quarks           & $\bm{30.2 \pm 0.1}$                   & $30.1 \pm 0.1$         & $25.9 \pm 0.3$   & $28.1 \pm 0.4$   & $28.2 \pm 0.4$        & $28.0 \pm 0.1$   \\
            W boson                & $70.4 \pm 0.1$                        & $65.8 \pm 0.1$         & $51.5 \pm 0.3$   & $70.2 \pm 0.3$   & $\bm{75.4 \pm 0.3}$   & $74.8 \pm 0.3$   \\
            Z boson                & $76.9 \pm 0.1$                        & $72.9 \pm 0.1$         & $66.4 \pm 0.2$   & $77.6 \pm 0.3$   & $\bm{80.9 \pm 0.3}$   & $80.5 \pm 0.2$   \\
            top quark              & $63.3 \pm 0.1$                        & $63.2 \pm 0.1$         & $60.3 \pm 0.3$   & $64.2 \pm 0.3$   & $\bm{64.7 \pm 0.5}$   & $63.6 \pm 0.4$   \\

            \hline
        \end{tabular}
    \end{adjustbox}
\end{table*}

The number of parameters and float-point operations (FLOP) for each model are listed in Table~\ref{tab:flops}. For MLP-Mixer models, these values are computed after fusing batch normalization into dense layers. The table clearly shows that MLP-Mixer models require significantly fewer parameters and FLOPs compared to JEDI-net models, while still achieving comparable or superior performance.

\stepcounter{footnote}
\footnotetext{The reported FLOPs may be overestimated, as certain matrix multiplications involving constant one-hot matrices are included in the FLOP calculations. However, these operations can be efficiently implemented as gather or lookup operations, significantly reducing computational complexity, as noted in ~\cite{jedi-fpga}.\label{fn:flop}}
\begin{table}[htb]
    \newcommand{\rr}[1]{\multirow{2}{*}{#1}}
    \centering
    \caption{The number of parameters and FLOPs for the MLP-Mixer and JEDI-net models. For MLP-Mixer models, values are computed after fusing batch normalization layers into dense layers.}
    \label{tab:flops}
    \begin{tabular}{l|lcc}
        \hline
        Model                                           & \multicolumn{1}{c}{Input} & Parameters & FLOPs                                 \\%& FLOPs/Parameter \\
        JEDI-net~\cite{jedi-net}                        & ${N_p}^{16}=100$          & {33,625}   & {116M\textsuperscript{\ref{fn:flop}}} \\%& \rr{3.4k}       \\
        JEDI-net w/ $\Sigma\mathcal{O}$~\cite{jedi-net} & ${N_p}^{16}=150$          & {8,767}    & {458M\textsuperscript{\ref{fn:flop}}} \\%& \rr{52.2k}      \\
        MLPM-fp                                         & ${N_p}^{16}=16$           & {2,465}    & {21.6k}                               \\%& \rr{8.8}        \\
        MLPM-fp                                         & ${N_p}^{16}=32$           & {3,265}    & {50.5k}                               \\%& \rr{15.5}       \\
        MLPM-fp                                         & ${N_p}^{16}=64$           & {6,401}    & {133k}                                \\%& \rr{20.8}       \\
        MLPM-fp                                         & ${N_p}^{16}=128$          & {18,817}   & {396k}                                \\%& \rr{21.0}       \\
        \hline
    \end{tabular}
\end{table}

\subsection{Quantized Models}
\label{sec:quantized}
In this section we evaluate the performance, resource consumption, and latency of the quantized models optimized for deployment on FPGAs.
For on-chip resource consumption, we consider the usage of Look-Up Tables (LUTs), digital signal processors (DSPs), Block RAMs (BRAMs), and flip-flops (FFs). All MLP-Mixer models presented in this work undergo HLS, logic synthesis, and place \& route using Vitis HLS 2023.2 and Vivado 2023.2. Resource consumption is reported post place \& route, ensuring accuracy in estimating the FPGA resource utilization. Timing convergence is verified and achieved for all models, unless explicitly stated otherwise. DA optimization using \texttt{da4ml} is applied for all models, and we use parallel I/O interfaces\footnote{Implemented using wire or register connections between layers, ensuring fully parallel data movement.} to minimize latency and maximize throughput. The same procedure is applied to the MLP models that were added to this study to ensure a fair comparison with simpler architectures.

The target FPGA is a Xilinx Virtex UltraScale+ XCVU9P (part number \texttt{xcvu9p-flga2104-2L-e}) operating at a 200 MHz clock frequency. To ensure bit-accurate consistency, we validate that the predictions from the quantized models obtained via the Python API precisely match those from the C-simulation of the HLS code across all samples in the test dataset. As a result, we do not differentiate between performance metrics obtained from Python and HLS simulations in subsequent evaluations. Latency and initiation interval\footnote{Number of blocking clock cycles required for processing each jet.} (II) are extracted from the HLS reports. Since the generated firmware does not include buffers or streaming interfaces, the reported latency and II can be considered accurate.

\begin{figure}[htb]
    \centering
    \subfigure[Accuracy vs. LUT usage]{
        \includegraphics[width=0.4\textwidth]{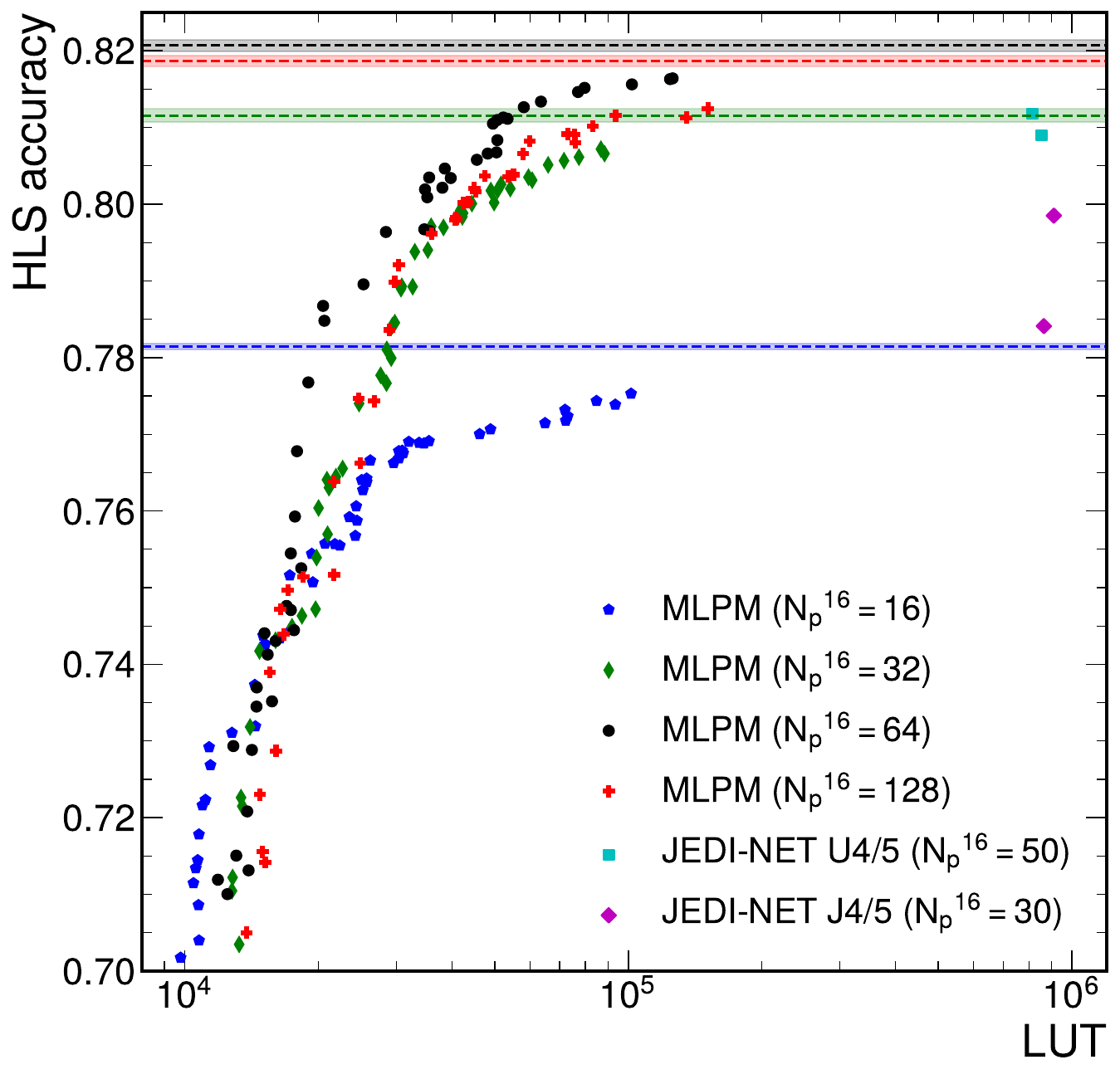}
        \label{fig:mlpm-16-lut}
    }
    \subfigure[Accuracy vs. Latency]{
        \includegraphics[width=0.4\textwidth]{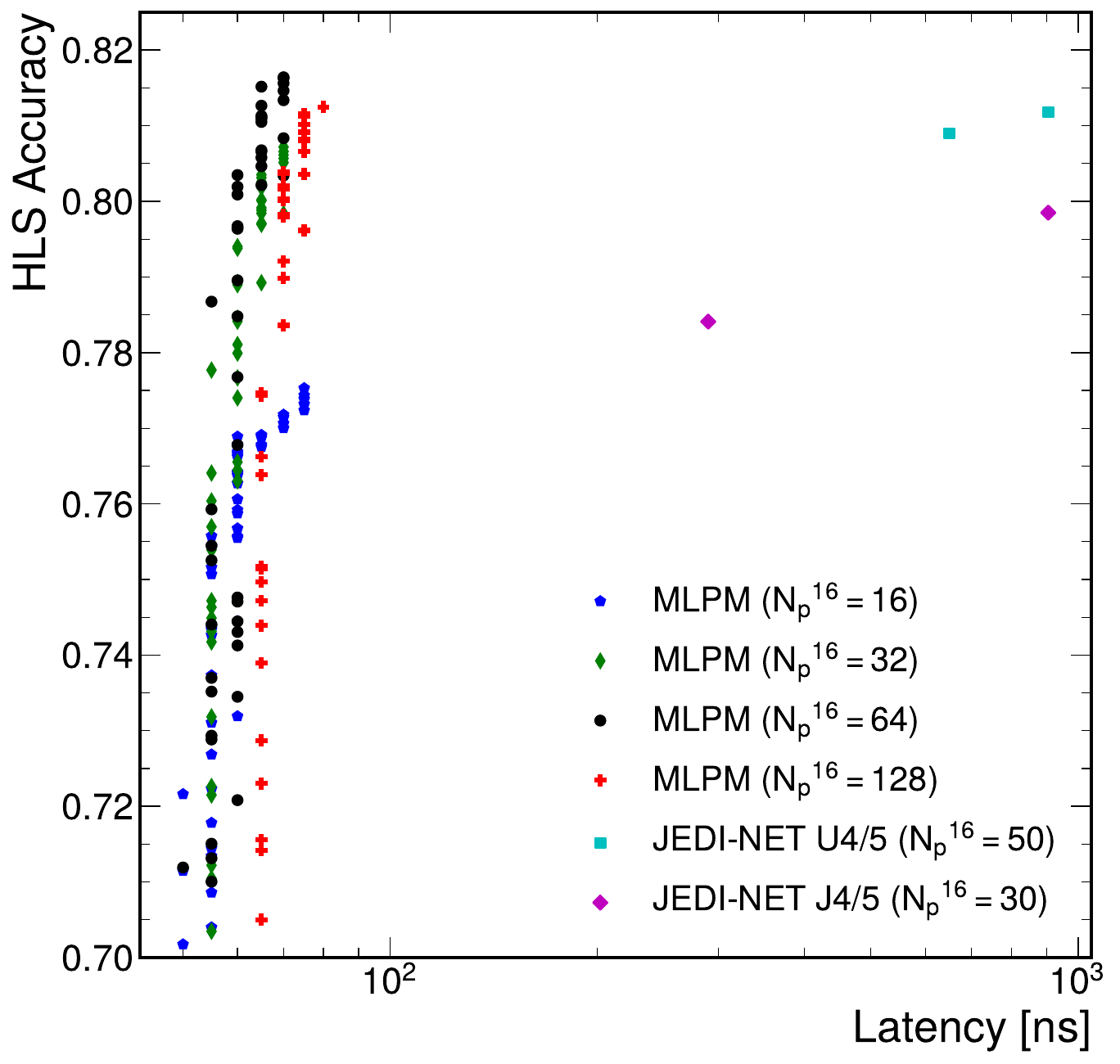}
        \label{fig:mlpm-16-latency}
    }
    \caption{Accuracy vs. LUT usage and latency for the quantized MLP-Mixer and JEDI-net models, each trained with 16 features per particle. Dashed lines indicate the accuracy of the corresponding full-precision models.}
\end{figure}

\begin{figure}[htb]
    \centering
    \subfigure[Accuracy vs. LUT usage]{
        \includegraphics[width=0.4\textwidth]{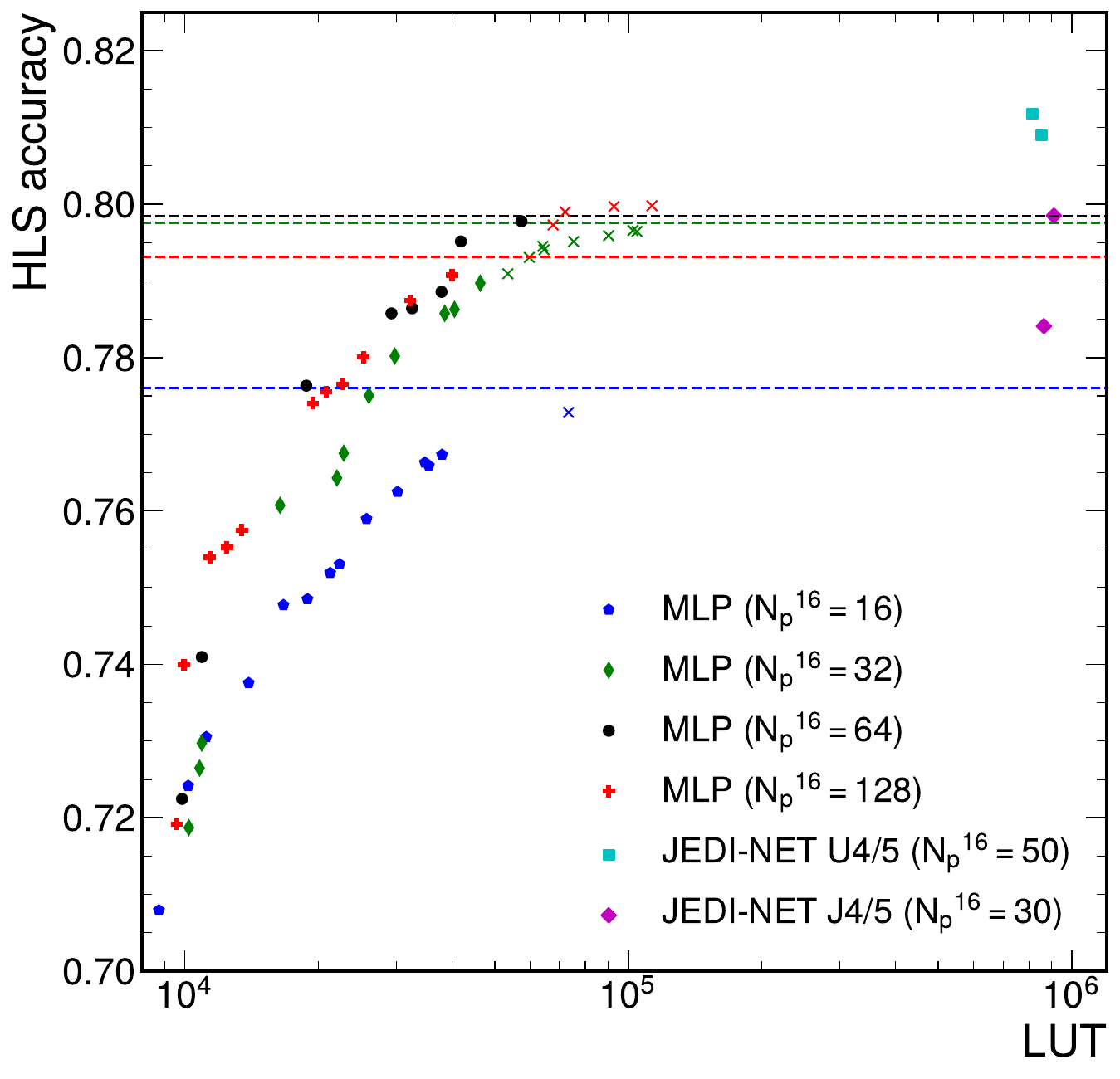}
        \label{fig:mlp-16-lut}
    }
    \subfigure[Accuracy vs. Latency]{
        \includegraphics[width=0.4\textwidth]{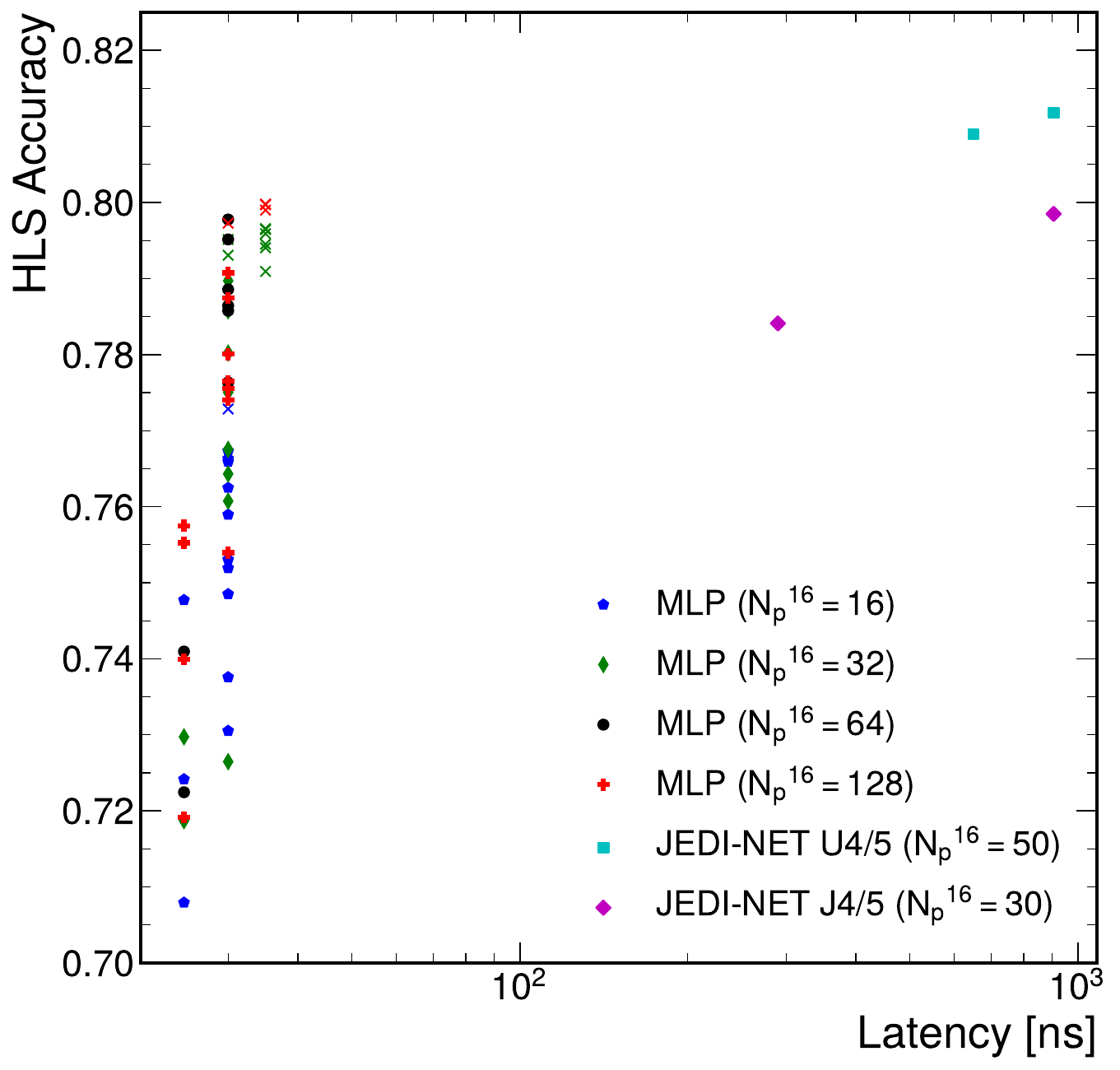}
        \label{fig:mlp-16-latency}
    }
    \caption{Accuracy vs. LUT usage and latency for the quantized MLP and JEDI-net models, each trained with 16 features per particle. Dashed lines indicate the accuracy of the corresponding full-precision models. Models marked with ``$\times$'' markers successfully underwent HDL synthesis but failed timing closure during the place \& route phase. These models' LUT usage are can still be used for reference, but their latencies reported are not accurate.}
\end{figure}

Figures~\ref{fig:mlpm-16-lut} and \ref{fig:mlpm-16-latency} illustrate the trade-off between accuracy, LUT usage, and latency for the quantized MLP-Mixer models, while Figure~\ref{fig:mlp-16-lut} and \ref{fig:mlp-16-latency} present the same comparison for the quantized MLP models. Since our DA optimization involves explicit unrolling, both MLP-Mixer and MLP models operate without using any DSP slices. The dashed lines in the figures indicate the accuracy of the corresponding full-precision models with the same number of input particles. For comparison, we also show the LUT usage, latency, and accuracy of the quantized JEDI-net models for ${N_p}^{16}$ = 30 and 50 from Ref.~\cite{jedi-fpga}.

The results demonstrate that MLP-Mixer models achieve a significantly better Pareto frontier between accuracy and resource consumption than JEDI-net models. All MLP-Mixer models operate within 100 ns latency, with the best-performing model achieving 81.64\% accuracy at 70 ns latency. Additionally, they deliver a two-order-of-magnitude higher throughput than JEDI-net models, while using fewer resources and achieving superior accuracy.

Although MLP models optimized with DA and HGQ achieve lower latency, their accuracy remains significantly lower than that of the MLP-Mixer models.

Table~\ref{tab:compare-16} presents a detailed comparison of selected quantized MLP-Mixer, MLP, and the JEDI-net models. In the table, $\mathrm{MLPM}_\mathrm{max}$ refers to the MLP-Mixer model with the highest accuracy for a given number of input particles, while $\mathrm{MLPM}_\mathrm{alt}$ represents an alternative model with slightly lower accuracy but significantly reduced resource consumption. Similarly, the best performing MLP model trained in this study is labeled as $\mathrm{MLP}_\mathrm{max}$. Across all configurations, MLP-Mixer models consistently outperform the JEDI-net models in terms of accuracy, LUT usage, latency, and throughput, achieving superior efficiency with a significant margin.

\begin{table*}[htb]
    \centering
    \caption{Comparison of quantized MLP-Mixer, MLP, and the JEDI-net models in terms of accuracy, resource consumption, and latency. $\mathrm{MLPM}_\mathrm{max}$ refers to the MLP-Mixer model with the highest accuracy for a given number of input particles, while $\mathrm{MLPM}_\mathrm{alt}$ represents an alternative model with slightly lower accuracy but significantly reduced resource usage. Similarly, $\mathrm{MLP}_\mathrm{max}$ denotes the best-performing MLP model trained in this study. All models are trained on the hls4ml jet tagging dataset with 16 features for input particle.}
    \label{tab:compare-16}
    \setlength\tabcolsep{2.5pt}
    \begin{adjustbox}{width=1.0\textwidth,center=\textwidth}
        \begin{tabular}{ll|ccccccc}
            \hline
            Model                        & ${N_p}^{16}$ & Accuracy (\%) & Latency      & DSP   & LUT (k) & FF (k) & BRAM & II  \\
            \hline
            JEDI-net U4~\cite{jedi-fpga} & 50           & 80.9          & 130 (650 ns) & 8,945 & 855     & 201    & 25   & 110 \\
            JEDI-net U5~\cite{jedi-fpga} & 50           & 81.2          & 181 (905 ns) & 8,986 & 815     & 189    & 37   & 150 \\

            JEDI-net J4~\cite{jedi-fpga} & 32           & 78.4          & 58 (290 ns)  & 8,776 & 865     & 138    & 37   & 30  \\
            JEDI-net J5~\cite{jedi-fpga} & 32           & 79.9          & 181 (905 ns) & 9,833 & 911     & 158    & 37   & 150 \\

            \hline

            MLPM\textsubscript{max}      & 16           & 77.5          & 15 (75 ns)   & 0     & 102     & 24     & 0    & 1   \\
            MLPM\textsubscript{max}      & 32           & 80.7          & 14 (70 ns)   & 0     & 87      & 22     & 0    & 1   \\
            MLPM\textsubscript{alt}      & 32           & 79.9          & 13 (65 ns)   & 0     & 42      & 11     & 0    & 1   \\
            MLPM\textsubscript{max}      & 64           & 81.6          & 14 (70 ns)   & 0     & 126     & 32     & 0    & 1   \\
            MLPM\textsubscript{alt}      & 64           & 81.3          & 13 (65 ns)   & 0     & 58      & 16     & 0    & 1   \\
            MLPM\textsubscript{max}      & 128          & 81.3          & 16 (80 ns)   & 0     & 151     & 42     & 0    & 1   \\

            \hline

            MLP\textsubscript{max}       & 16           & 77.3          & 6 (30 ns)    & 0     & 73      & 18     & 0    & 1   \\
            MLP\textsubscript{max}       & 32           & 79.7          & 7 (35 ns)    & 0     & 103     & 23     & 0    & 1   \\
            MLP\textsubscript{max}       & 64           & 79.8          & 6 (30 ns)    & 0     & 57      & 15     & 0    & 1   \\
            MLP\textsubscript{max}       & 128          & 80.0          & 7 (35 ns)    & 0     & 113     & 29     & 0    & 1   \\
            \hline
        \end{tabular}
    \end{adjustbox}
\end{table*}

While the models saturate at different accuracy levels, we observe that the trade-off curves for the MLP-Mixer and MLP models with varying number of input particles remain mostly consistent -- particularly in the accuracy region below 78\% -- with the exception of the 16-particles variants. This suggests that relying on parameter count or FLOPs as a proxy for resource consumption in quantized models can be misleading, as these metrics scale strongly with input size, as shown in Table~\ref{tab:flops}. To provide further insights, Table~\ref{tab:bitwidths} presents the bitwidths for weights and activations in the MLP-Mixer models, as defined in \cite{hgq}.
The data clearly shows that as the number of input particles increases, both the bitwidths and $(1-$sparsity$)$ of the weights and activations decrease, effectively compensating for the otherwise higher resource consumption.

\begin{table*}[htb]
    \centering
    \caption{Average bitwidths (Avg. BW), average bitwidths of non-zero elements (Avg. BW (non-zero)), and sparsity for weights and activations of the $\mathrm{MLPM}_\mathrm{max}$ models. Activation sparsity refers to the static sparsity after quantization, which is independent of input data during inference.}
    \setlength\tabcolsep{5pt}
    \label{tab:bitwidths}
    \begin{adjustbox}{width=1.0\textwidth,center=\textwidth}
        \begin{tabular}{ll|ccc|ccc}
            \hline
            \multirow{3}{*}{Model}  & \multirow{3}{*}{${N_p}^{16}$} &                          & Weights    &                           &                          & Activations &                           \\
            \cline{3-8}
                                    &                               & \multirow{2}{*}{Avg. BW} & Avg. BW    & \multirow{2}{*}{Sparsity} & \multirow{2}{*}{Avg. BW} & Avg. BW     & \multirow{2}{*}{Sparsity} \\
                                    &                               &                          & (non-zero) &                           &                          & (non-zero)  &                           \\
            \hline
            MLPM\textsubscript{max} & 16                            & 1.93                     & 2.52       & 0.23                      & 4.80                     & 6.34        & 0.24                      \\
            MLPM\textsubscript{max} & 32                            & 1.55                     & 2.51       & 0.38                      & 3.27                     & 5.66        & 0.42                      \\
            MLPM\textsubscript{alt} & 32                            & 0.60                     & 2.27       & 0.73                      & 1.78                     & 4.82        & 0.63                      \\
            MLPM\textsubscript{max} & 64                            & 1.06                     & 2.44       & 0.57                      & 2.55                     & 5.19        & 0.51                      \\
            MLPM\textsubscript{alt} & 64                            & 0.58                     & 2.29       & 0.75                      & 1.67                     & 4.81        & 0.65                      \\
            MLPM\textsubscript{max} & 128                           & 0.57                     & 2.27       & 0.75                      & 1.78                     & 4.76        & 0.62                      \\
            \hline
        \end{tabular}
    \end{adjustbox}
\end{table*}

\begin{figure*}
    \centering
    \subfigure[RoC for MLP-Mixers with ${N_p}^{16}=16$]{
        \includegraphics[width=0.4\textwidth]{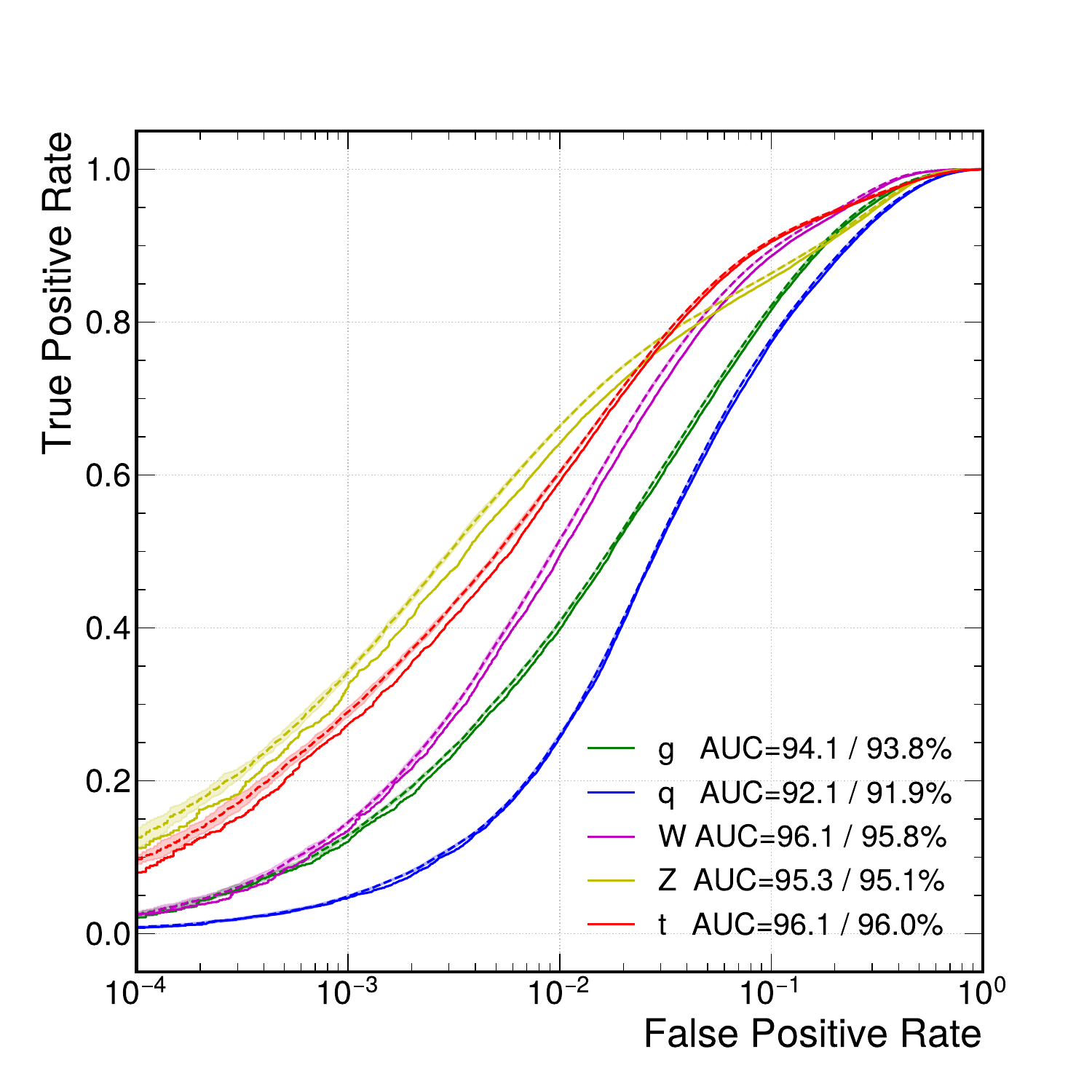}
    }
    \subfigure[RoC for MLP-Mixers with ${N_p}^{16}=32$]{
        \includegraphics[width=0.4\textwidth]{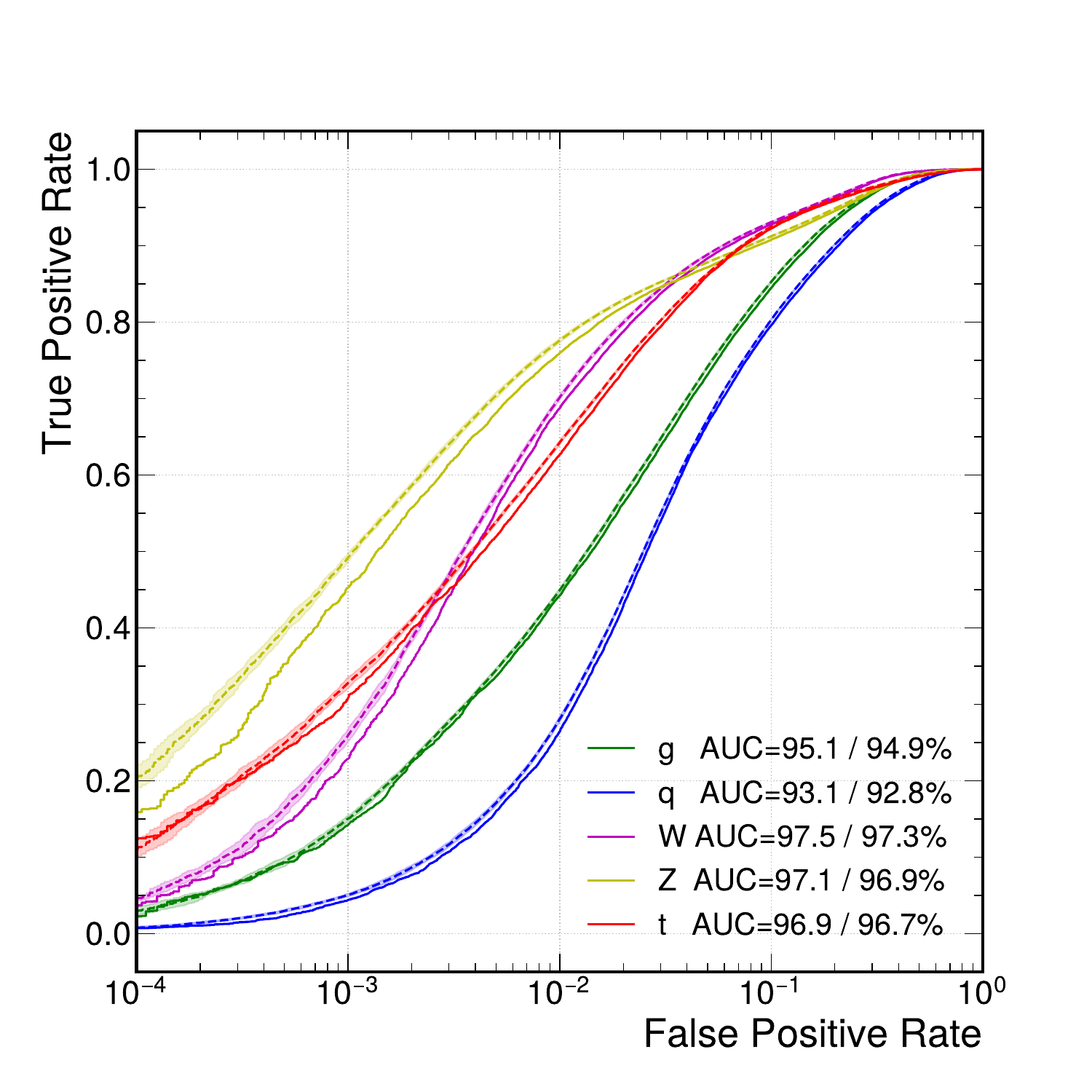}
    }\\
    \subfigure[RoC for MLP-Mixers with ${N_p}^{16}=64$]{
        \includegraphics[width=0.4\textwidth]{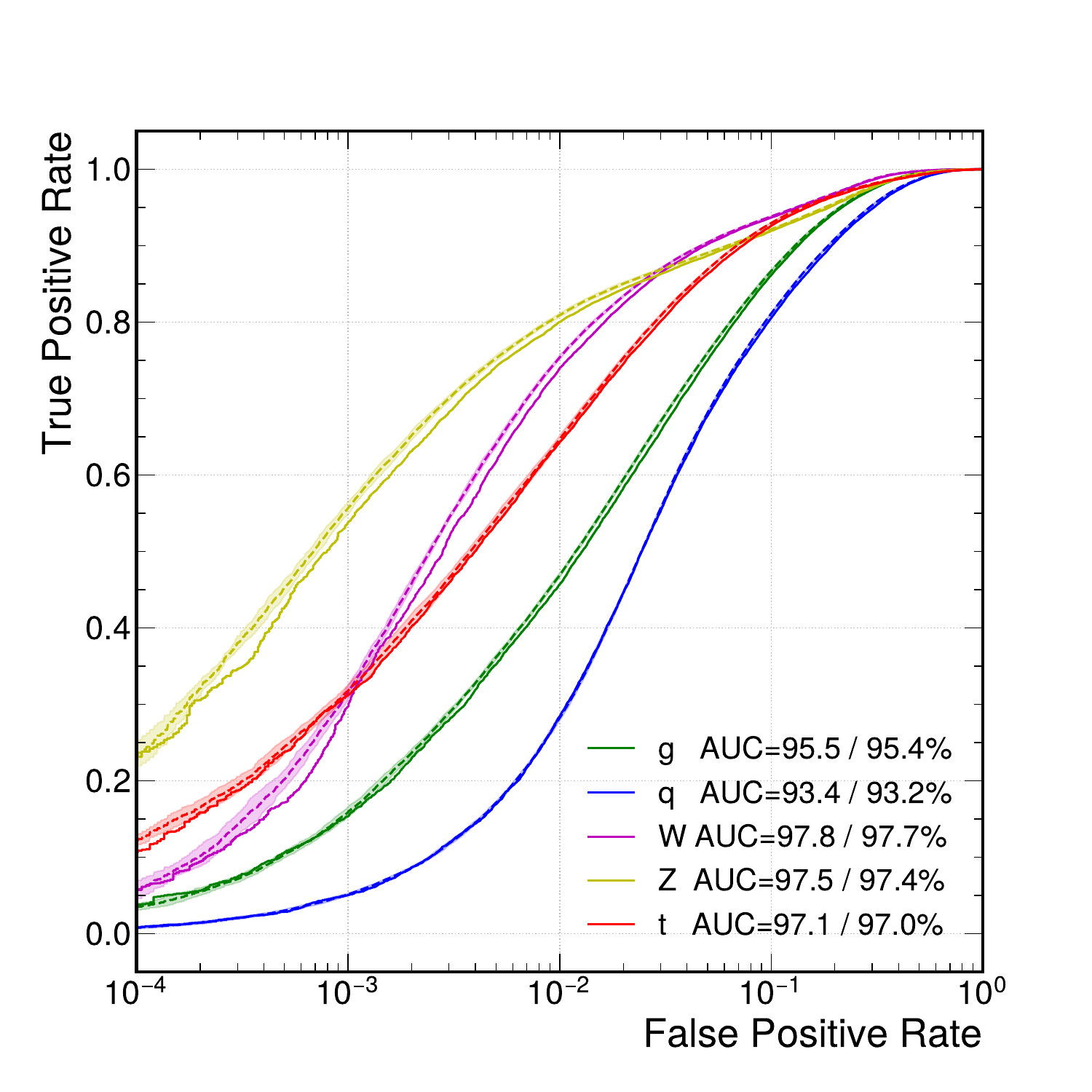}
    }
    \subfigure[RoC for MLP-Mixers with ${N_p}^{16}=128$]{
        \includegraphics[width=0.4\textwidth]{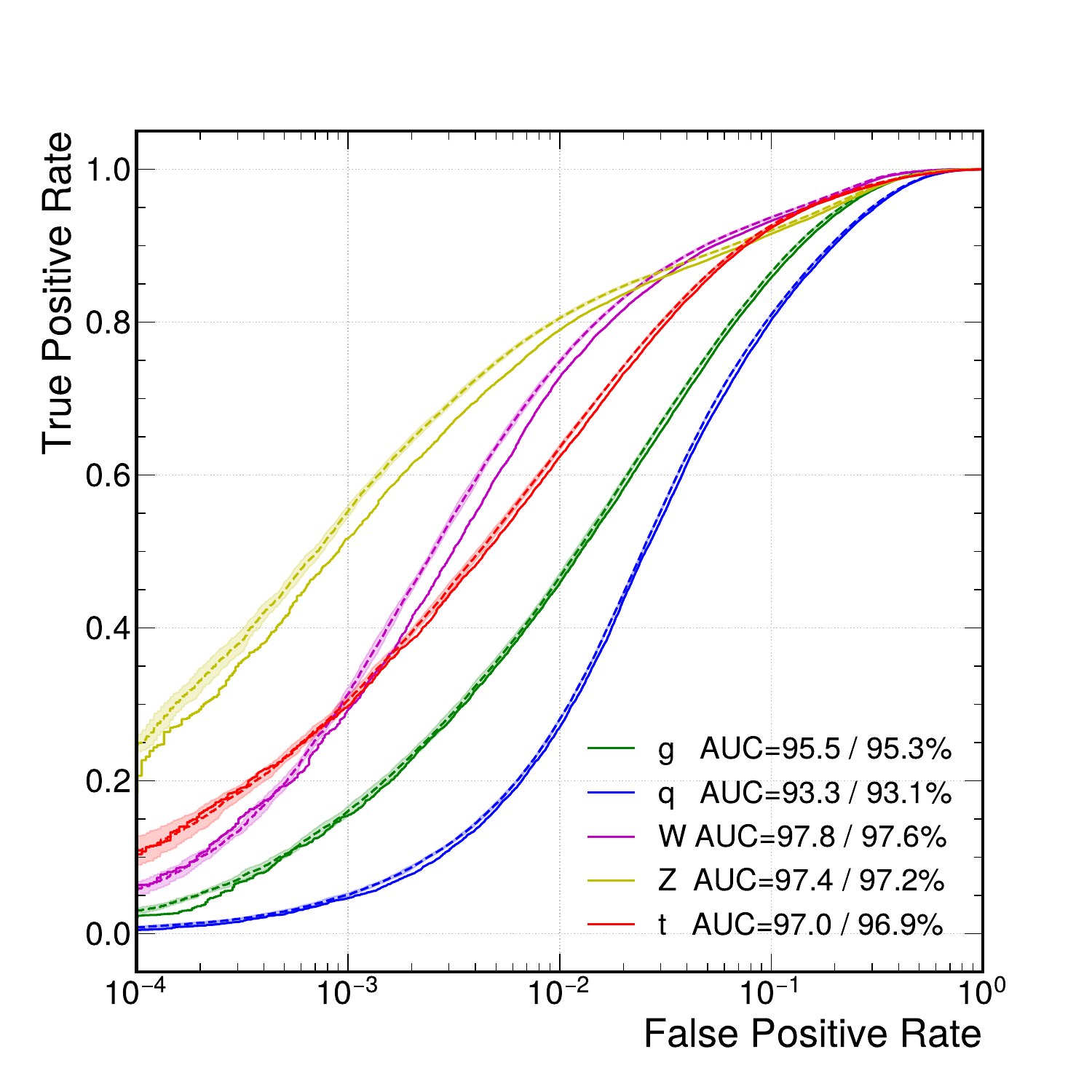}
    }
    \caption{Receiver Operating Characteristic (RoC) curves of the $\mathrm{MLPM}_\mathrm{max}$ models (solid lines) compared to their full precision counterpart (dashed lines) trained with 16, 32, 64, and 128 input particles. The Area Under the Curve (AUC) for each jet category is provided in the legend, formatted as "full-precision / quantized".}
    \label{fig:roc-16}
\end{figure*}

Finally, the Receiver Operating Characteristic (RoC) curves for the full-precision and best performing quantized MLP-Mixer models with 16, 32, 64, and 128 input particles are shown in Figure~\ref{fig:roc-16}. In all cases, the quantized models achieve comparable, but slightly degraded performance compared to the full-precision models.

\begin{figure*}[htb]
    \centering
    \includegraphics[width=0.9\textwidth]{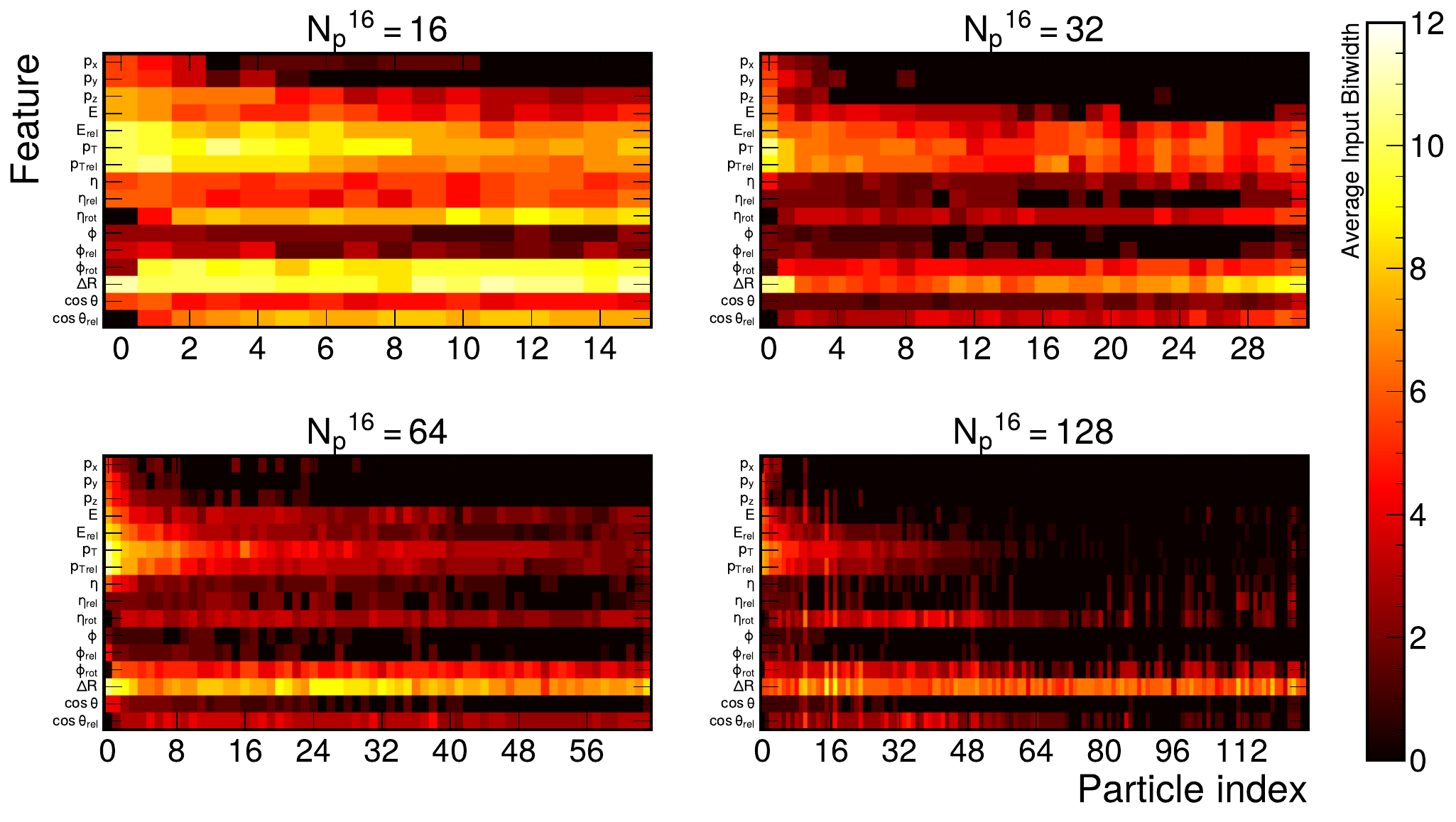}
    \caption{Average input bitwidths for the best-performing quantized MLP-Mixer models with 16, 32, 64, and 128 input particles. The reported values are computed by summing the bitwidths at the input to MLP1 and one leg of the addition operation before MLP3, then dividing by two.}
    \label{fig:bw}
\end{figure*}

\subsection{Feature Importance}

In an HGQ-trained model, the bitwidth of activations can serve as a proxy for feature importance. When quantizing a value drawn from a smooth distribution, the quantization-induced change can be interpreted as additive noise affecting the original value. Since the integer bitwidth is adjusted to accommodate the maximum activation value, the quantization noise for a given value $x$ is approximately proportional to $\max{x} \cdot 2^{-\text{bitwidth}}$. This suggests that bitwidth allocation reflects the model's tolerance for additive noise on a given feature when trading-off with resource usage: features assigned higher bitwidths are considered more critical, as the model minimizes quantization errors for them. However, this interpretation is less reliable when input values originate from highly discrete or narrowly distributed distributions compared to the magnitude of the quantization noises.

Figure~\ref{fig:bw} presents the average input bitwidths for the four best-performing quantized MLP-Mixer models. These are obtained by adding and dividing by two the bitwidths at two key locations: (i) at the input to MLP1 and (ii) before the addition operation in MLP3 (refer to Figure \ref{fig:model}).
From Figure~\ref{fig:bw}, we observe that the models allocate more bits to particles carrying higher energy, particularly for the features $E$, $E_\text{rel}$, $p_T$, and ${p_T}_\text{rel}$, which are assigned higher bitwidths for the first $\sim40$ particles. In contrast, $\eta_\text{rot}$, $\phi_\text{rot}$, $\Delta R$ receive higher bitwidths across all particles, indicating their persistent importance in jet tagging. This variation in bitwidth allocation suggests that MLP-Mixer models actively discard less relevant features to optimize resource usage. Furthermore, it reinforces the advantage of breaking permutation invariance in jet tagging tasks. If the model were forced to maintain permutation invariance, bitwidths would need to remain constant across particle indices for each feature, leading to substantial resource overhead. This highlights the efficiency gains achieved by allowing the model to prioritize key features dynamically.

\subsection{Particle-wise Feature prioritization}
\label{sec:feature-prioritization}
In this section, we perform an ablation study to evaluate the impact of particle-wise feature prioritization on the efficiency of the MLP-Mixer models: different bitwidths are only assigned to the different channels, but all particles will share the same bitwidth for each channel. This is in contrast to the heterogeneous quantization used in Section~\ref{sec:quantized}, where each particle can have a different bitwidths, allowing the model to prioritize features based on their importance for each particle, as shown in Figure~\ref{fig:bw}.

\begin{figure}[htb]
    \centering
    \includegraphics[width=0.4\textwidth]{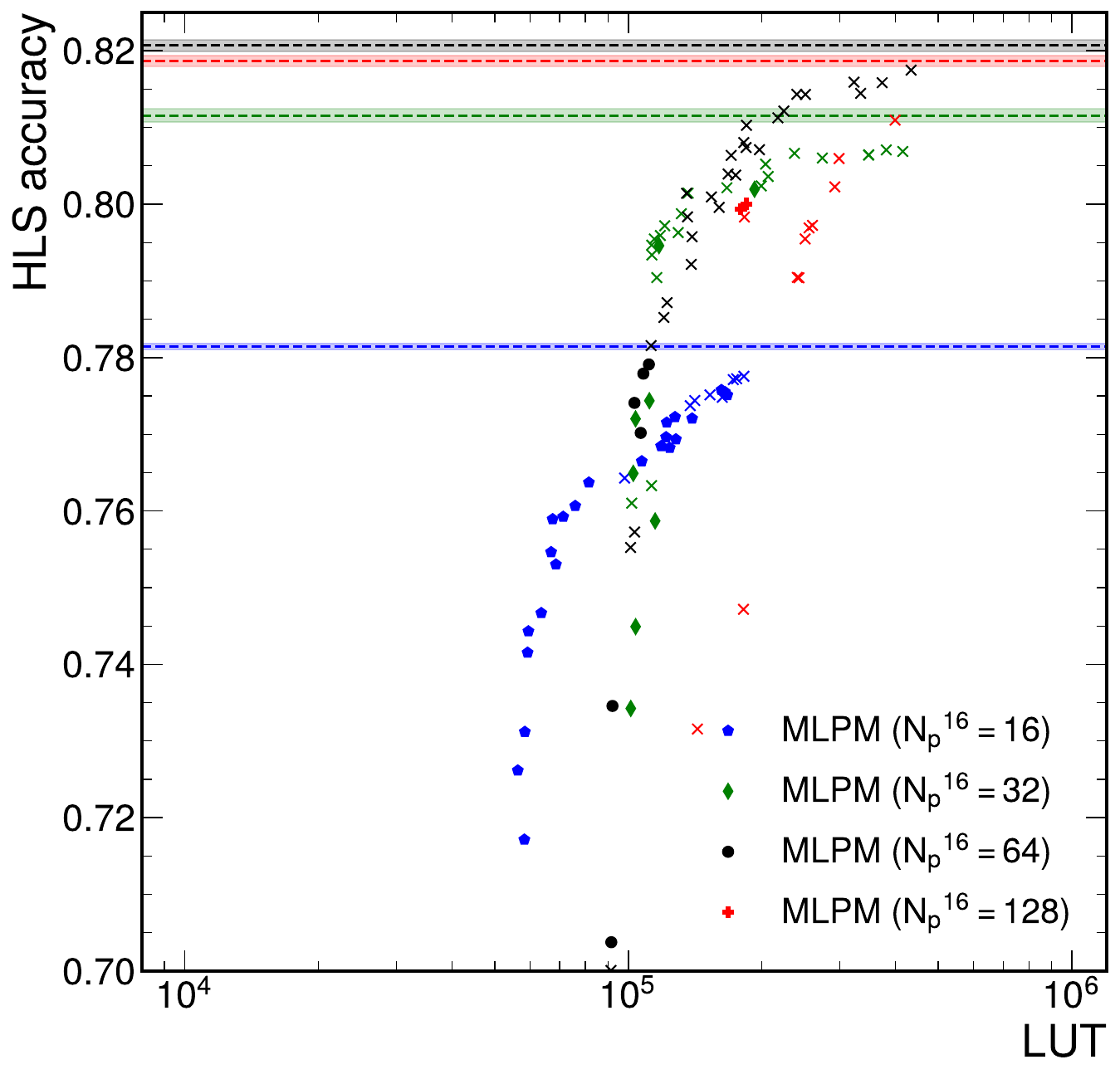}
    \label{fig:mlpm-16-lut-uniform}
    \caption{Accuracy vs. LUT usage and latency for the quantized MLP-Mixer without particle-wise heterogeneous quantization, each trained with 16 features per particle. Dashed lines indicate the accuracy of the corresponding full-precision models. Models marked with ``$\times$'' markers successfully underwent HDL synthesis but failed timing closure during the place \& route phase. These models' LUT usage are can still be used for reference, but their latencies reported are not accurate.}
\end{figure}

All other hyperparameters and other configuration remain unchanged from Section~\ref{sec:quantized}. The results are presented in Figure~\ref{fig:mlpm-16-lut-uniform} for the accuracy vs. LUT usage trade-off. Latency vs. accuracy trade-off is not shown as the majority of models do not meet timing, and the latency reported this way will not be accurate. As the models are fully and pipelined, adding more pipeline stages to close timing would not incur significant LUT overhead, and therefore the models LUT usage with failed timing are still useful for comparison.

Comparing to the results shown in Figure~\ref{fig:mlpm-16-lut}, the models with uniform bitwidths for each input particle require significant more LUTs to achieve the same accuracy as the models without it. The difference is particularly pronounced for the models with 64 and 128 input particles, where the models with uniform particle-wise quantization require one order of magnitude more LUTs to achieve the same accuracy as the models with fully heterogeneous quantization. As uniform bitwdiths for each particle is a prerequisite for permutation invariance neural networks, these results strongly suggests that permutation invariance could lead to major resource overhead.

\subsection{Quantized model with $p_T$, $\eta$ and $\phi$ inputs}

\begin{figure}[htb]
    \centering

    \subfigure[Accuracy vs. LUT usage]{
        \includegraphics[width=0.4\textwidth]{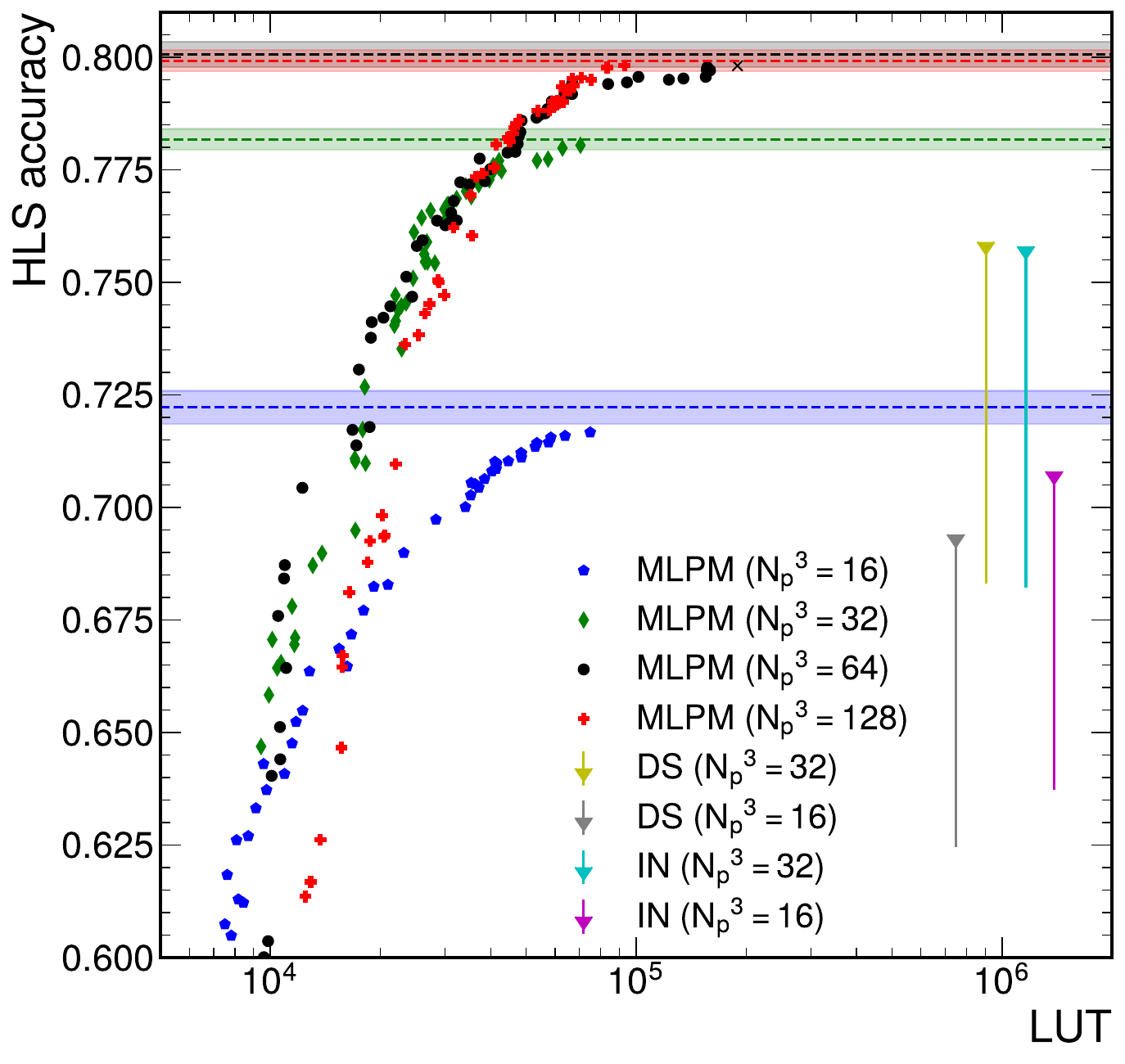}
        \label{fig:mlpm-3-lut}
    }
    \subfigure[Accuracy vs. Latency]{
        \includegraphics[width=0.4\textwidth]{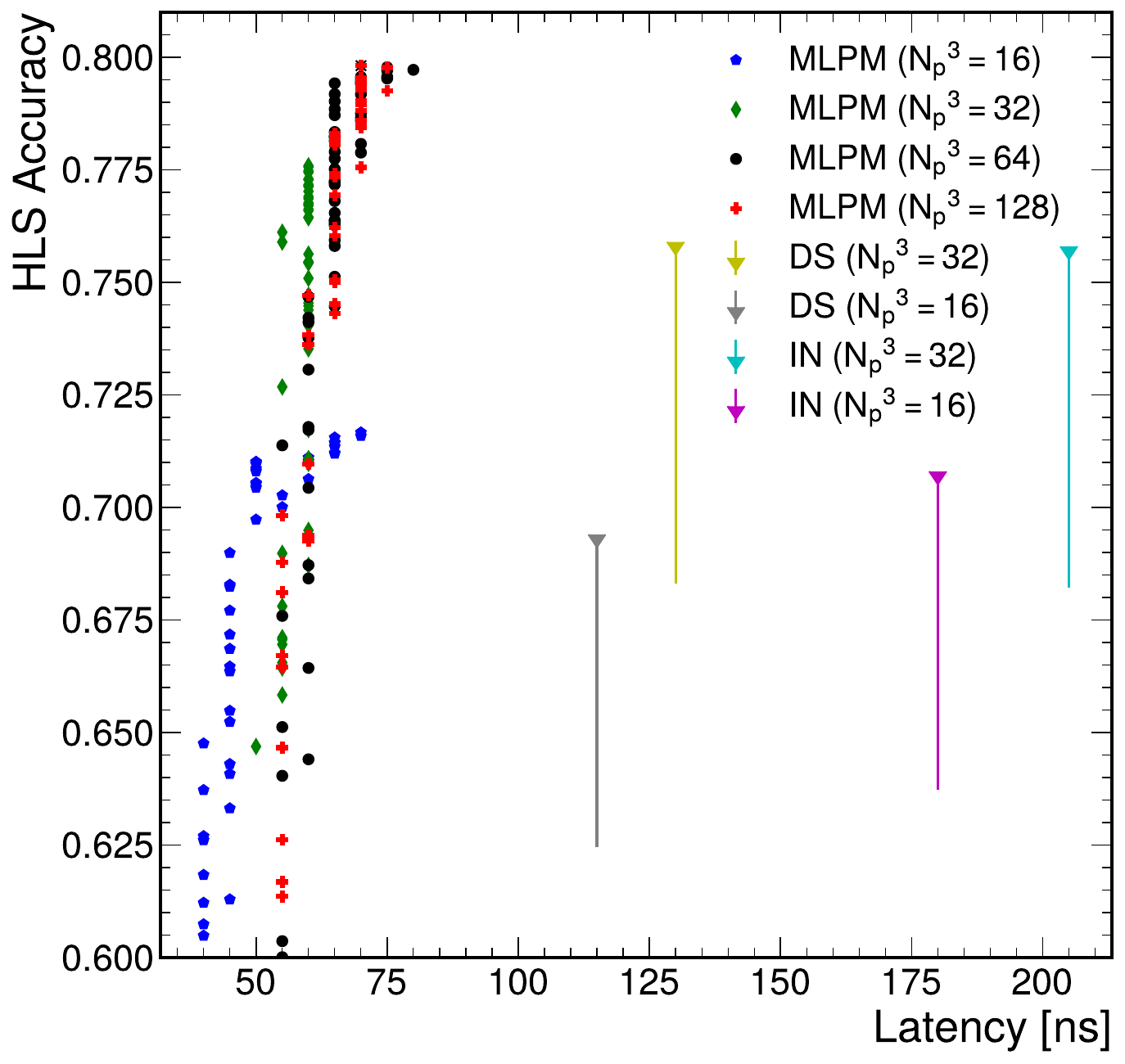}
        \label{fig:mlpm-3-latency}
    }
    \caption{Accuracy vs. LUT usage (left) and latency (right) of the quantized MLP-Mixer, Interaction Network (IN)~\cite{ds-fpga}, and Deep Sets (DS)~\cite{ds-fpga} models trained using only $p_T$, $\eta$, and $\phi$ as input features per particle. Dashed lines indicate the accuracy of the corresponding full-precision models with the same number of input particles. Models marked with a ``$\times$'' successfully underwent HDL synthesis but failed timing closure during the place \& route phase. These models' LUT usage are can still be used for reference, but their latencies reported are not accurate. The error bars for the IN and DS models reflect ambiguity in the reported accuracy in the original work~\cite{ds-fpga}.}
\end{figure}

\begin{figure}[htb]
    \centering

    \subfigure[Accuracy vs. LUT usage]{
        \includegraphics[width=0.4\textwidth]{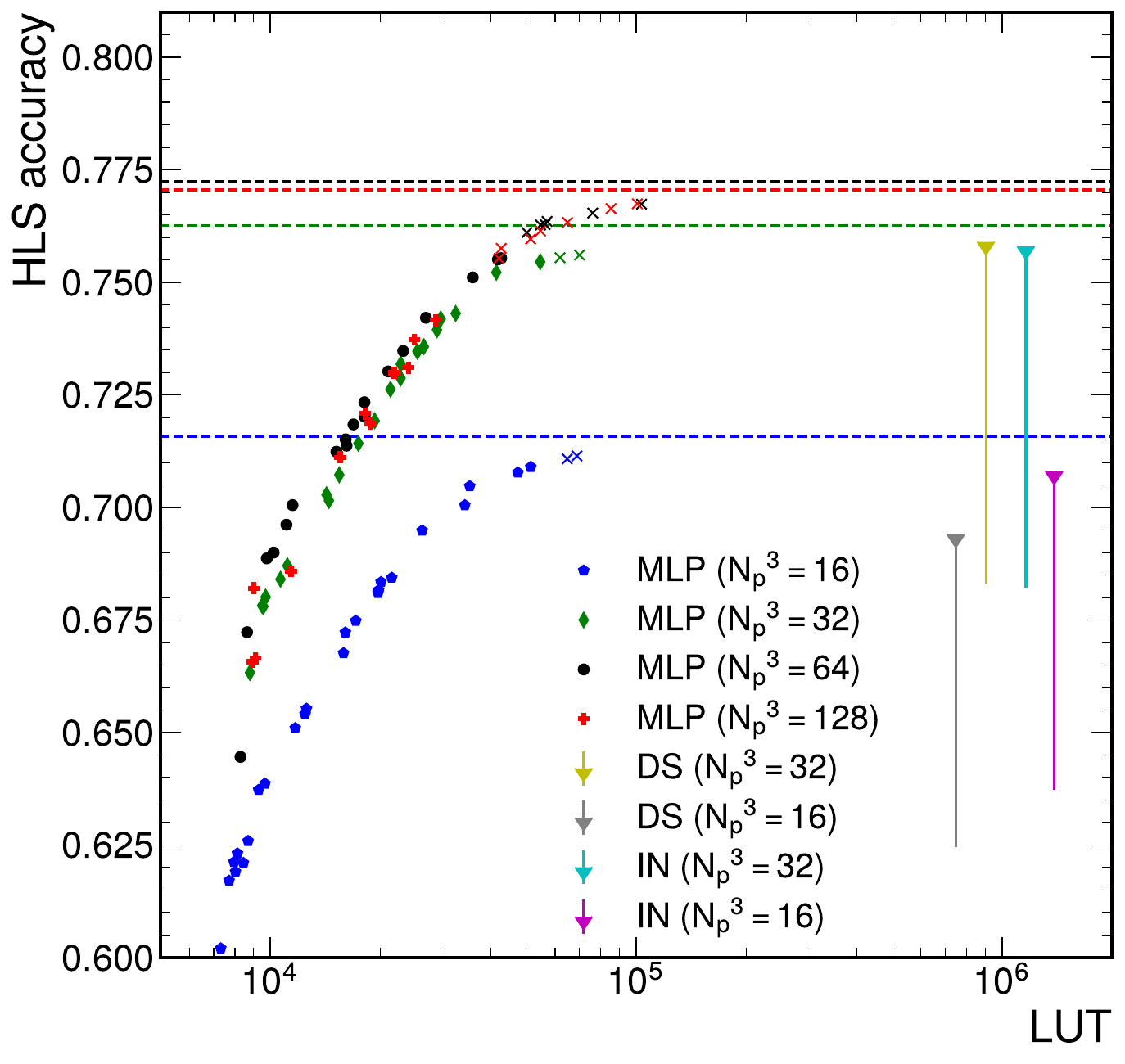}
        \label{fig:mlp-3-lut}
    }
    \subfigure[Accuracy vs. Latency]{
        \includegraphics[width=0.4\textwidth]{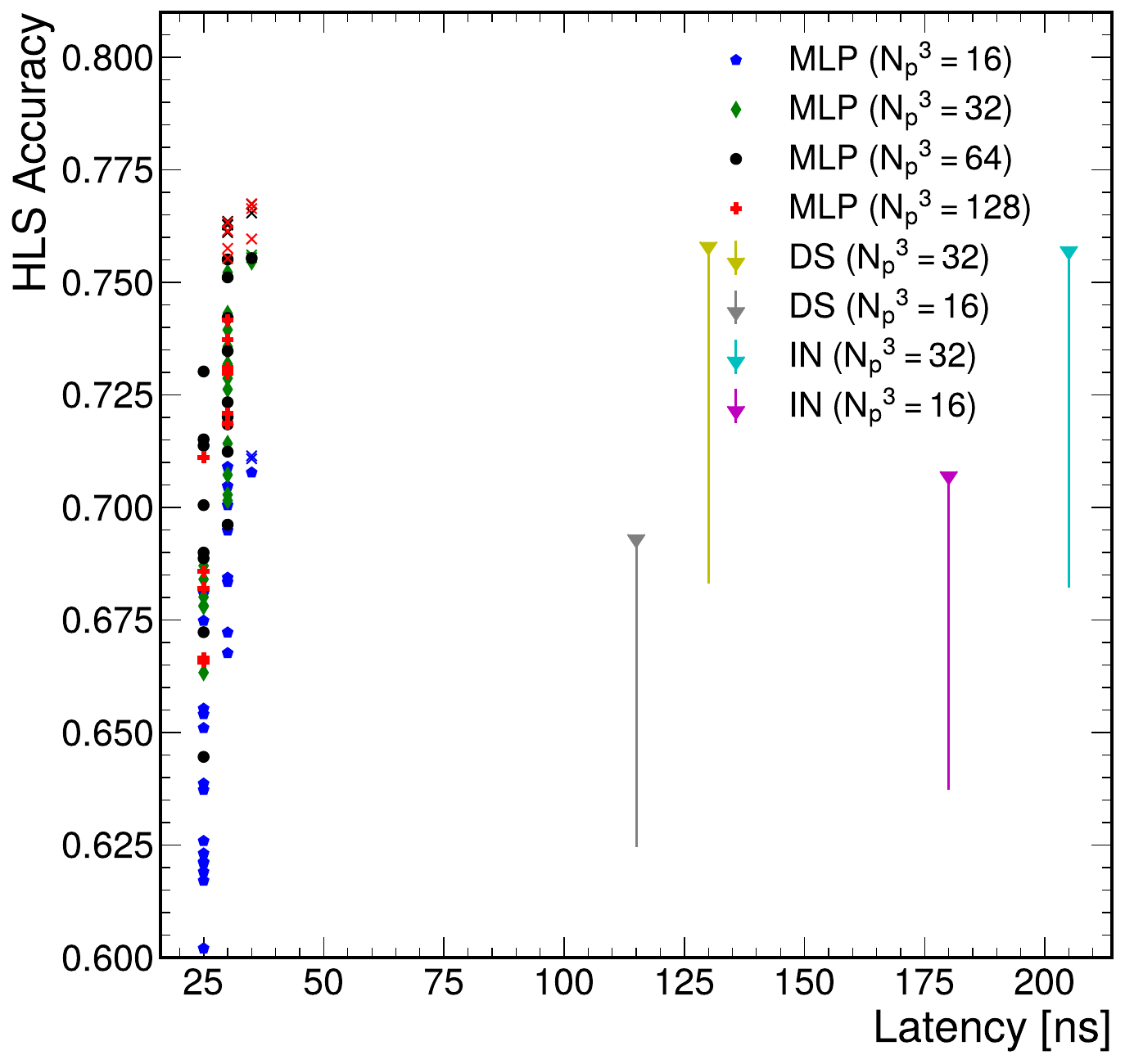}
        \label{fig:mlp-3-latency}
    }

    \caption{Accuracy vs. LUT usage (left) and latency (right) of the quantized MLP, Interaction Network (IN)~\cite{ds-fpga}, and Deep Sets (DS)~\cite{ds-fpga} models trained with using only $p_T$, $\eta$, and $\phi$ as input features per particle. Dashed lines indicate the accuracy of the corresponding full-precision models with the same number of input particles. Models marked with a ``$\times$'' successfully underwent HDL synthesis but failed timing closure during the place \& route phase. These models' LUT usage are can still be used for reference, but their latencies reported are not accurate. The error bars for the IN and DS models reflect ambiguity in the reported accuracy in the original work~\cite{ds-fpga}.}
\end{figure}

\begin{table*}[htb]
    \centering
    \caption{Performance, latency, and resource consumption of the quantized MLP-Mixer, MLP, DS and IN models described by the accuracy and resource consumption. All models are trained on the hls4ml jet tagging dataset using only $p_T$, $\eta$, and $\phi$ as input features per particle. Only particles with $p_T\ge2$ GeV are included in the training and evaluation. The intervals for accuracies for the DS and IN models are due to the ambiguity in the original work~\cite{ds-fpga}, where the author did not report model accuracies after quantization and firmware generation.}
    \label{tab:compare-3}
    \setlength\tabcolsep{5pt}
    \begin{tabular}{ll|ccccccc}
        \hline
        Model                   & ${N_p}^{3}$ & Accuracy (\%) & Latency     & DSP   & LUT (k) & FF (k) & BRAM & II \\
        \hline
        DS~\cite{ds-fpga}       & 16          & $62.5 - 69.4$ & 23 (115 ns) & 555   & 747     & 239    & 0    & 3  \\
        IN~\cite{ds-fpga}       & 16          & $63.7 - 70.8$ & 36 (180 ns) & 5,362 & 1,388   & 594    & 0    & 3  \\
        DS~\cite{ds-fpga}       & 32          & $68.3 - 75.9$ & 26 (130 ns) & 434   & 903     & 359    & 0    & 2  \\
        IN~\cite{ds-fpga}       & 32          & $68.2 - 75.8$ & 41 (205 ns) & 2,120 & 1,162   & 761    & 0    & 3  \\

        \hline
        MLPM\textsubscript{max} & 16          & 71.7          & 14 (70 ns)  & 0     & 75      & 17     & 0    & 1  \\
        MLPM\textsubscript{alt} & 16          & 70.8          & 10 (50 ns)  & 0     & 40      & 8.4    & 0    & 1  \\
        MLPM\textsubscript{max} & 32          & 78.0          & 13 (65 ns)  & 0     & 63      & 15     & 0    & 1  \\
        MLPM\textsubscript{alt} & 32          & 76.1          & 11 (55 ns)  & 0     & 25      & 7.1    & 0    & 1  \\
        MLPM\textsubscript{max} & 64          & 79.7          & 15 (75 ns)  & 0     & 159     & 36     & 0    & 1  \\
        MLPM\textsubscript{alt} & 64          & 75.9          & 13 (65 ns)  & 0     & 26      & 7.9    & 0    & 1  \\
        MLPM\textsubscript{max} & 128         & 79.8          & 15 (75 ns)  & 0     & 83      & 21     & 0    & 1  \\

        \hline

        MLP\textsubscript{max}  & 16          & 71.1          & 7 (35 ns)   & 0     & 69      & 14     & 0    & 1  \\
        MLP\textsubscript{max}  & 32          & 75.6          & 7 (35 ns)   & 0     & 70      & 13     & 0    & 1  \\
        MLP\textsubscript{max}  & 64          & 75.5          & 7 (35 ns)   & 0     & 43      & 8.7    & 0    & 1  \\
        MLP\textsubscript{max}  & 128         & 76.7          & 7 (35 ns)   & 0     & 101     & 19     & 0    & 1  \\

        \hline
    \end{tabular}
\end{table*}

\begin{figure*}[htb]
    \centering
    \subfigure[RoC for MLP-Mixers with ${N_p}^{3}=16$]{
        \includegraphics[width=0.4\textwidth]{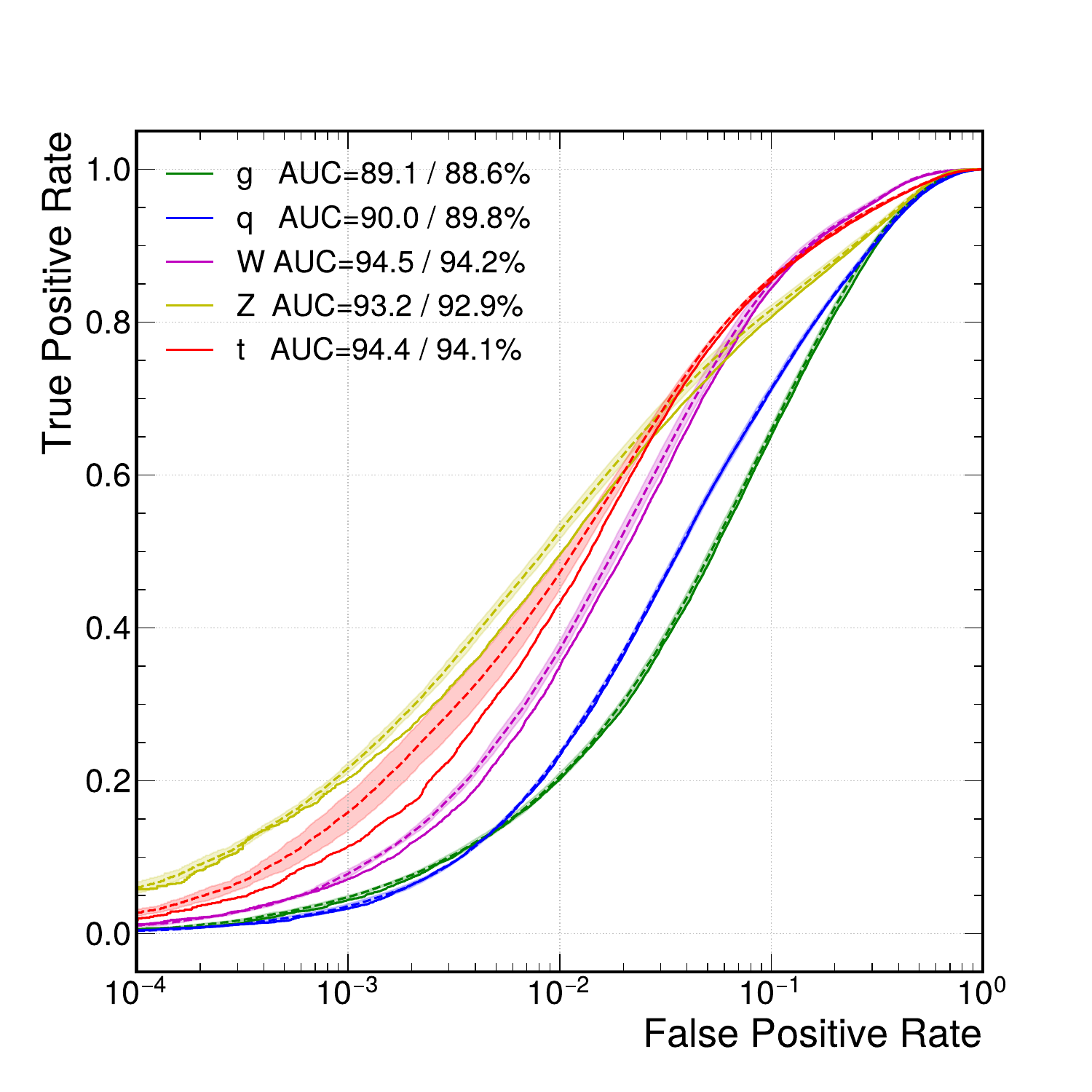}
    }
    \subfigure[RoC for MLP-Mixers with ${N_p}^{3}=32$]{
        \includegraphics[width=0.4\textwidth]{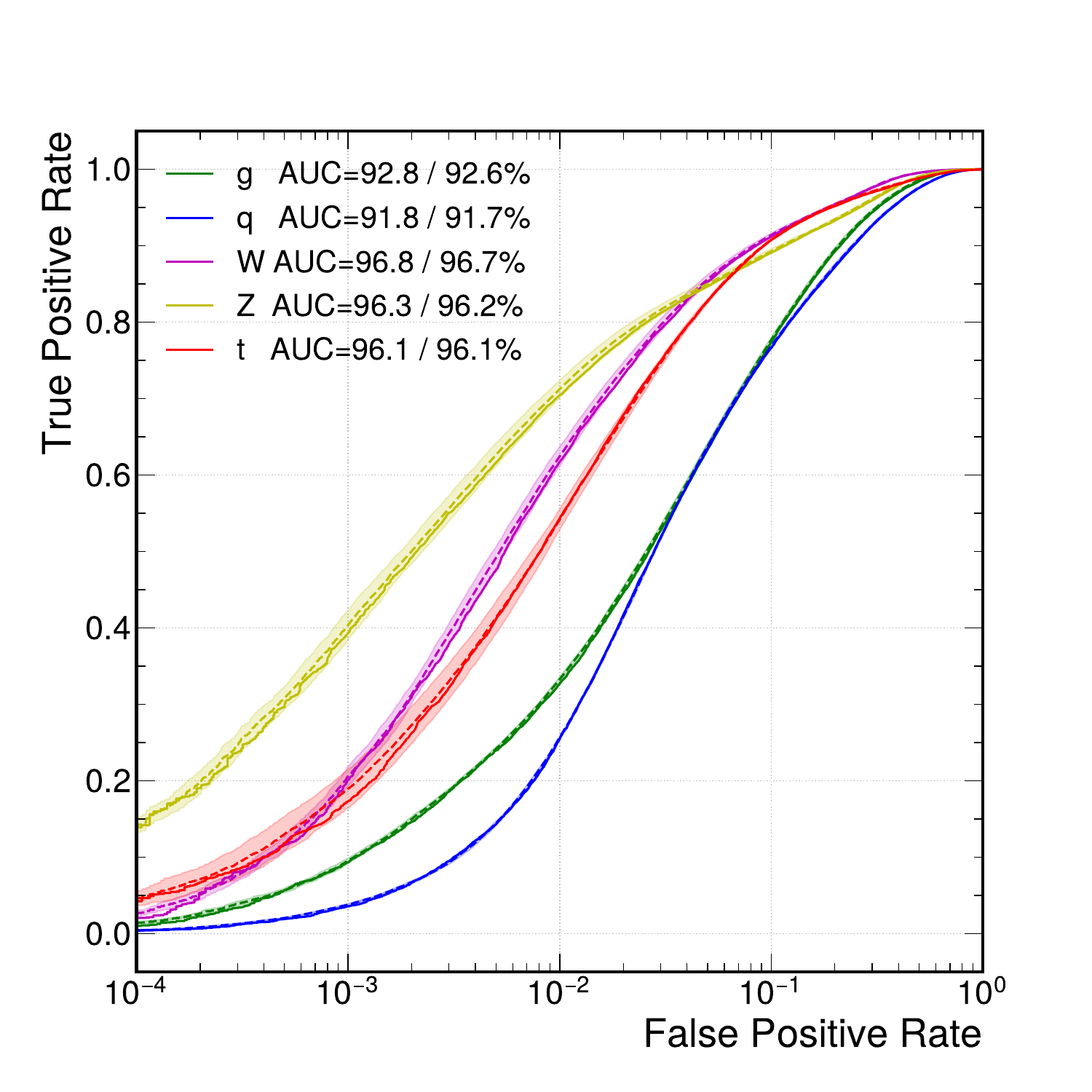}
    }\\
    \subfigure[RoC for MLP-Mixers with ${N_p}^{3}=64$]{
        \includegraphics[width=0.4\textwidth]{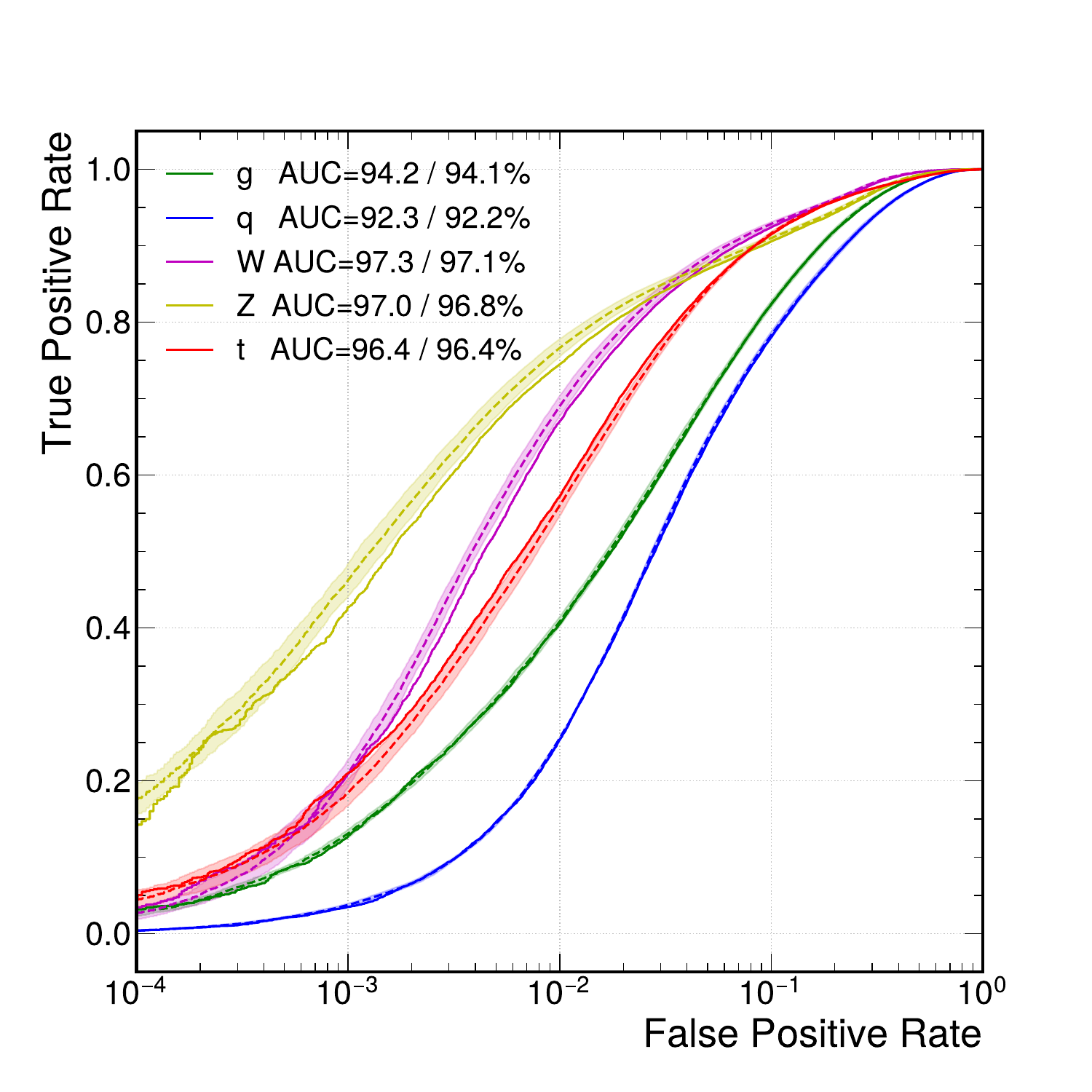}
    }
    \subfigure[RoC for MLP-Mixers with ${N_p}^{3}=128$]{
        \includegraphics[width=0.4\textwidth]{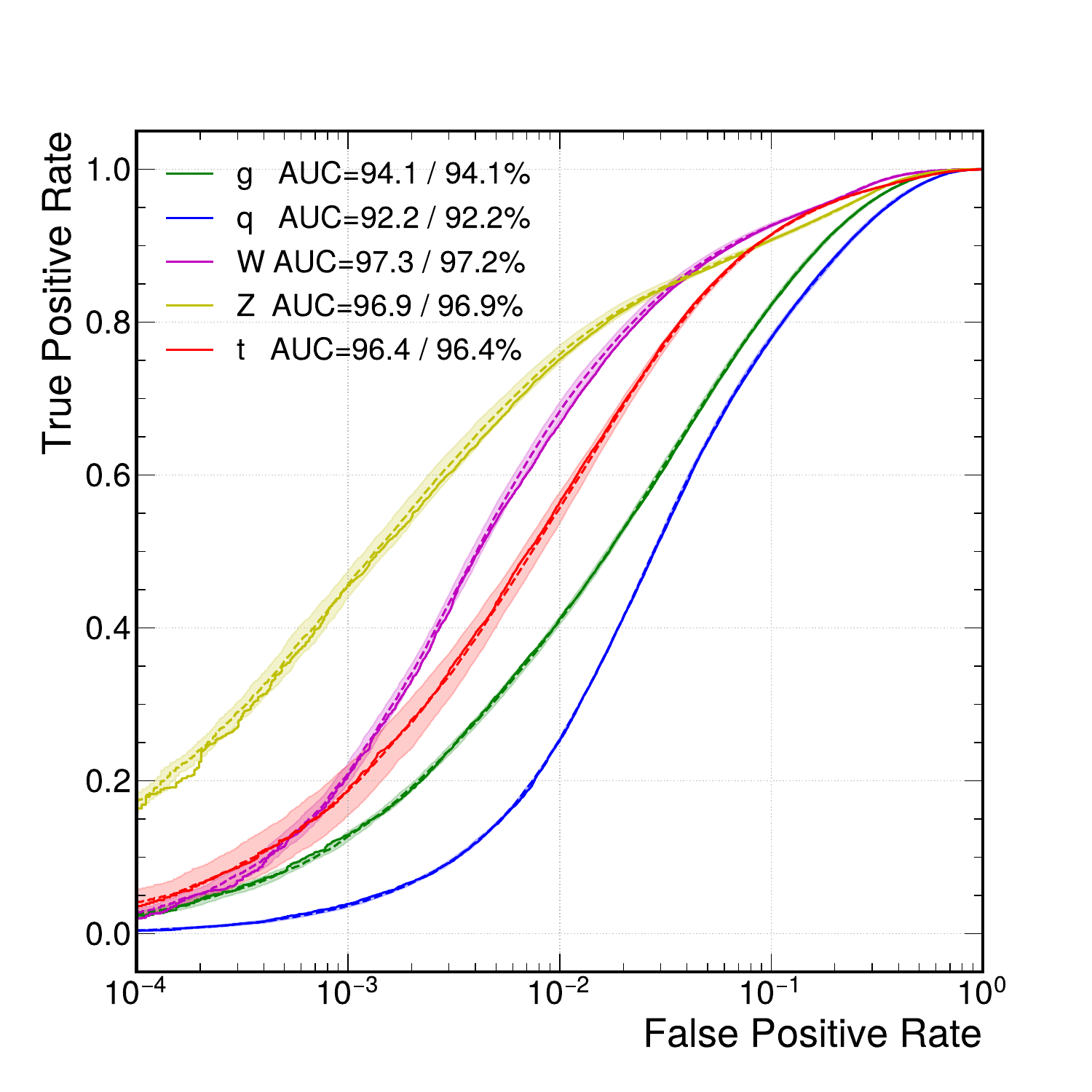}
    }
    \caption{Receiver Operating Characteristic (RoC) curves of the MLPM\textsubscript{max} models (solid lines) compared to their full precision counterpart (dashed lines) trained with 16, 32, 64, and 128 input particles and using only $p_T$, $\eta$, and $\phi$ as input features per particle. The Area Under the Curve (AUC) for each jet category is provided in the legend, formatted as ``full-precision / quantized''.}
    \label{fig:roc-3}
\end{figure*}

In this section, we evaluate the performance of the quantized MLP-Mixer models trained using only $p_T$, $\eta$, and $\phi$ for each particle, rather than the full 16-feature set described in Table~\ref{tab:features}. This configuration reflects the minimal set of features may be provided by the hardware trigger systems, where all kinematic features beyond momentum may not always be available at L1T. In addition to restricting input features, we apply a $p_T\ge 2$ GeV selection to each particle as performed in Ref.~\cite{ds-fpga}, to better approximate detector behavior. Apart from adjusting the layer sizes to accommodate the reduced feature set (as shown in Figure~\ref{fig:model} with $n=3$), all other hyperparameters remain unchanged. Training and evaluation procedures follow the same methodology used for models trained with 16 features per particle, as detailed in Sections~\ref{sec:full_precision} and~\ref{sec:quantized}.

The performance of the quantized MLP-Mixer models trained with particle features $p_T$, $\eta$, and $\phi$ is shown in Figures~\ref{fig:mlpm-3-lut} and \ref{fig:mlpm-3-latency}. The corresponding results for the quantized MLP models we trained are shown in Figures~\ref{fig:mlp-3-lut} and \ref{fig:mlp-3-latency}. These models are compared with the 8-bits quantized Interaction Network (IN) and Deep Sets (DS) models proposed in Ref.~\cite{ds-fpga} where they were trained with the same inputs. For the IN and DS models, the error bars on the accuracy reflect ambiguities in the reported performance in the original work, where only full-precision accuracy was provided. The accuracy of their quantized counterparts was estimated to be at least 90\% of their full-precision versions. As shown in the figures, the quantized MLP-Mixer models consistently outperform the IN and DS models in both accuracy and resource consumption.

A detailed comparison of selected MLP-Mixer, MLP, IN, and DS models is provided in Table~\ref{tab:compare-3}. The notation follows Table~\ref{tab:compare-16}.
We exclude MLP-Mixer and MLP models that failed timing closure in the place \& route phase from the comparison.
Depending on the chosen MLP-Mixer model, we demonstrate that MLP-Mixers can outperform IN and DS models by orders of magnitude in resource efficiency while maintaining comparable or superior accuracy.
For instance, the MLPM\textsubscript{alt} (${N_p}^3=32$) model achieves the following improvements over the best IN and DS models for each metric:
\begin{itemize}
    \item 0.2\% higher accuracy than the full precision models
    \item 97\% lower LUT usage
    \item 97\% lower FF usage
    \item 100\% lower DSP usage
    \item 52\% lower latency
    \item 100\% higher throughput
\end{itemize}
Similar to models trained with 16 particles as input, the MLP models optimized with DA and HGQ achieve lower latency but are limited in accuracy compared to MLP-Mixer models. Nevertheless, our results demonstrate that when properly optimized, even MLP models can surpass permutation-invariant architectures (IN and DS models) in both resource efficiency and accuracy, as shown in Figure~\ref{fig:mlp-3-lut} and \ref{fig:mlp-3-latency}.

Finally, Figure~\ref{fig:roc-3} presents the ROC curves for the full-precision and best-performing quantized MLP-Mixer models trained with particle features $p_T$, $\eta$, and $\phi$. The quantized models achieve comparable, albeit slightly lower, performance relative to their full-precision counterparts.

\clearpage
\section{Conclusions}
In this work, we demonstrated that MLP-Mixer is a highly effective architecture for jet classification tasks using particle-level feature inputs. Our results show that MLP-Mixer models outperform JEDI-net models in accuracy, while requiring significantly fewer FLOPs, leading to faster inference times on CPUs and GPUs. Furthermore, we demonstrated that by leveraging HGQ and \texttt{da4ml}, MLP-Mixer models can be efficiently compressed and deployed on FPGAs. These models achieve a superior Pareto frontier between accuracy and resource consumption compared to previous models, while also delivering the highest accuracy, throughput, and lowest latency among the models evaluated.

Although MLP-Mixer models are not permutation invariant, we showed that they can dynamically discard less relevant features to optimize resource usage. Additionally, our analysis revealed that the models selectively allocate more bits to particles with higher $p_T$, suggesting that enforcing permutation invariance could introduce unnecessary resource overhead. Based on these findings, we argue that breaking permutation invariance can be beneficial for the jet classification task, particularly when the input data is known to be sorted by a physically meaningful quantity.

For future works, we identify a few improvements could be made over this work, and hope they could be useful for the community:
\begin{itemize}
    \item the kernels for the particle-wise mixers (MLP2 and MLP4 in Figure~\ref{fig:model}) could be regularized for mitigating overfitting. In the models discussed in this work, only the ${N_p}^{16}=128$ models showed slight signs of overfitting in the full-precision and a few least quantized models. However, this could be more pronounced with larger models or datasets.
    \item In this work, for the feature-wise mixers (MLP1 and MLP3), the kernel applied on each particle is exactly identical. It may be beneficial to allow for heterogeneous bitwidths the kernels applied to each particle to further optimize resource usage. This would require modifications in the quantization frameworks, but the authors believe that the benefits would be worth the effort.
    \item For distributed arithmetic implementation, this work uses HLS for implementing the firmware. Targeting specific FPGA architecture, direct mapping the logic to LUTs and other logic primitives may be beneficial. For instance, using a compressor trees for summing over the partial outputs of the distributed arithmetic would lead to further reduction in resource usage.
\end{itemize}

\nocite{gnu_parallel}

\section*{Data availability statement}
\label{sec:data_avail}
The data and software required to reproduce this work can be found at \href{https://doi.org/10.5281/zenodo.3602260}{https://doi.org/10.5281/zenodo.3602260}~\cite{hls4ml-dataset} and \href{https://github.com/calad0i/HGQ-demos/tree/master/jet_classifier_large}{https://github.com/calad0i/HGQ-demos/tree/master/jet\_classifier\_large}.

\section*{Acknowledgements}
C.S. acknowledges partially supported by the NSF ACCESS Grant number PHY240298. C.S. and M.S. acknowledge partial support from the U.S. Department of Energy (DOE), Office of Science, Office of High Energy Physics grant DE-SC0011925. J.N., M.S., and C.S. are partially supported by the U.S. Department of Energy (DOE), Office of Science, Office of High Energy Physics ``Designing efficient edge AI with physics phenomena'' Project ({DE-FOA-0002705}). J.N. is partially supported by the AI2050 program at Schmidt Futures (Grant G-23-64934).

\section*{References}
\bibliographystyle{iopart-num}
\bibliography{bibliography}

\end{document}